\documentstyle[12pt]{article}
\begin{document}
\begin{titlepage}
\title{
\vspace{2 cm}
Heavy Baryon Chiral Perturbation Theory with Light Deltas\footnote{
Revised Version of {\tt [TRI-PP-97-21,ZU-TH 25/97]}; to appear in Journal of
Physics {\bf G}.}}
\vspace{2 cm}
\author{Thomas R. Hemmert$^{a,b}$\footnote{email: th.hemmert@fz-juelich.de},
Barry R.
Holstein$^{a,c}$\footnote{email: holstein@phast.umass.edu} \\
        and Joachim Kambor$^d$\footnote{email: kambor@sonne.physik.unizh.ch}
\\[1.5 cm]
{\small $^a$ Forschungszentrum J{\" u}lich, IKP (Th.)} \\
{\small D-52425 J{\" u}lich, Germany} \\
{\small $^b$ TRIUMF Theory Group, 4004 Wesbrook Mall}\\
{\small Vancouver, BC, Canada V6T 2A3}\\
{\small $^c$ Department of Physics and Astronomy, University of Massachusetts}
\\
{\small Amherst, MA 01003, USA} \\
{\small $^d$ Institut f\"ur Theoretische Physik, Universit\"{a}t Z\"{u}rich}\\
{\small CH-8057
Z\"{u}rich, Switzerland}}
%\nodate
\maketitle

\begin{abstract}
We demonstrate how heavy mass methods, previously applied to
chiral perturbation theory calculations involving the interactions of
nucleons and pions, can be generalized to include interactions with the
$\Delta$(1232) in a systematic formalism which we call the
``small scale expansion''.
\end{abstract}

\end{titlepage}

\tableofcontents
\newpage

\section{Introduction}

The subject of chiral perturbation theory has become an important one
in contemporary physics, as it represents a procedure by which to make rigorous
contact between experimental measurements and the QCD Lagrangian which
is thought to generate all hadronic interactions.\cite{a}  The way in which
this is done is to make use of the underlying chiral symmetry
of QCD in order to construct effective Lagrangians in terms of hadrons
which retain this symmetry.  Of course, in the real world chiral
symmetry is broken by quark mass effects but these are assumed to be
small and therefore treatable perturbatively.  This program was begun
in the 1960's but stalled when it was not recognized how to deal with the
infinite number of such effective Lagrangians which can be written
down.\cite{b}  In addition such theories are not renormalizable and this also
served as a detriment to further development.
Renewed interest, however, was generated in 1979 when a
seminal paper by Weinberg demonstrated how to solve both problems.\cite{c}  The
issue of non-renormalizability was shown to be a red herring.  True,
a full renormalization of such effective field theories would involve
introduction of an infinite number of possible counterterm Lagrangians
in order to cancel loop divergences.  However, Weinberg showed that
provided that one stays at energy,momentum low compared to the chiral
scale\cite{d}---$E,p<\Lambda_\chi\sim$ 1 GeV---then a
consistent renormalization scheme is possible and only a
finite number of possible structures and their associated counterterms
must be dealt with.  These counterterm contributions not only remove
loop divergences but also include finite pieces whose size can be
determined empirically.  Such terms encode the contributions from
higher energy sectors of the theory, whose form need not be given
explicitly.  In addition a consistent power counting scheme
can be developed.  The point here is that the lowest order effective
chiral Lagrangian is well-known to be of order $p^2$, meaning that it
involves either two powers of energy-momentum or of the
pseudoscalar mass, while it is easy to see that a one loop diagram
generates terms of order $p^4$.  Similarly it is straightforward to
characterize any such contributions by its associated power of ``momentum''
and the program of chiral perturbation theory becomes feasible.

The program was actually carried out by Gasser and Leutwyler who
developed a successful formalism to one loop---${\cal O}(p^4)$---in
the sector of Goldstone boson ($\pi,K,\eta$) interactions.\cite{e}  Their
papers
stimulated considerable work in this area, and consistent two-loop
calculations are now being performed.\cite{f}  This work has been reviewed in
a number of places and it is not necessary to repeat it here.\cite{g,chpt97}

A second arena of activity in this field has been in the low energy
properties of the light baryons. In this case, however, things are not so
straightforward.  The problem is that, in addition to the ``small''
dimensionful parameters of energy-momentum
and pseudoscalar mass, there exists also a ``large'' dimensionful
number---the baryon mass $M_B$---which is comparable in size to the
chiral scale itself, thus rendering the idea of a consistent
perturbative treatment doubtful. Indeed Gasser, Sainio and Svarc \cite{GSS88}
calculated $\pi -N$ scattering in a relativistic framework and showed that
no consistent power counting in analogy to the meson-sector exists.
Nevertheless several calculations in the relativistic framework were performed
in the 1980s, {\it e.g.} \cite{BSW,Krause,BGKM}.
However, by generalizing new developments in the field of heavy quark effective
theory (HQET) \cite{IW}, Jenkins and Manohar
showed how ``heavy baryon'' methods could be used to eliminate the
large mass term by going to the
non-relativistic limit. \cite{JM91}
Subsequently Bernard et al. showed that a consistent power
counting formulation of heavy baryon chiral perturbation theory (HBChPT)
was possible \cite{BKKM92} and
did extensive calculations which fully developed the power of such
methods for the low energy $\pi-N$ sector. A detailed review of
this work is given in ref. \cite{j}.
More recently, important contributions to the renormalization of the theory
were made by Ecker \cite{Eck94} and by Ecker and Moijzis \cite{EM96,EM97} for
the case
of SU(2) and Mei{\ss}ner and M{\" u}ller \cite{guido} for the case of SU(3).
Furthermore, for an overview of
the rapidly evolving field of effective chiral lagrangians for the
nucleon-nucleon interaction and applications to few body systems we refer
to \cite{rho,vanK97}.

Despite the excellent work done by these groups, a number of problems
remain.  One is that the convergence properties of the perturbative
series may require inclusion of an unfortunately large number of
terms, especially in the SU(3) version of HBChPT.  This is not yet clear,
however, and is the subject of current study. \cite{strange}
A second difficulty is the way in which the contribution of
resonant baryon states is handled.  In the systematic works mentioned above
\cite{BKKM92,j,Eck94,EM96,EM97,guido} it is assumed that such
states are very heavy compared to the nucleon.  In this case they can
be integrated out and replaced by a finite piece of a counterterm
contribution\footnote{There are plenty of ``chiral calculations'' in the
literature which employ effective chiral lagrangians with explicit resonance
degrees
of freedom, for example see \cite{rho,BSS93} and references therein. In this
work we do
not comment specifically on these calculations but are
only concerned with a {\it systematic extension} of the highly developed
$SU$(2) HBChPT
$\pi N$ formalism of refs.\cite{BKKM92,j,Eck94,EM96,EM97} to $\pi -N-\Delta$
interactions.}.  However, while this may be a reasonable scheme for heavier
resonances such as the Roper and higher states, it is of questionable
validity for the case of the $\Delta$(1232), which lies a mere 300 MeV
above the nucleon ground state and which couples very strongly to
the $\pi-N$ sector.  In fact, because of this strong coupling $\Delta$(1232)
contributions begin, in general, quite soon above threshold in those
channels wherein such effects are possible.  This suggests that
instead of including the effects of the resonance by simple
counterterms, it would be useful to include $\Delta$(1232) as an explicit
degree of freedom in the effective lagrangian. This has also been advocated by
Jenkins and Manohar \cite{JM91b} and their collaborators \cite{BSS93}.

Over the past few years \cite{HHK96a} we have developed a consistent
chiral power counting scheme--the so called ``small scale expansion''--which
builds upon the systematic HBChPT formalism of
refs.\cite{BKKM92,j,Eck94,EM96,EM97}
and in addition allows for explicit nucleon and delta degrees of freedom to be
treated
simultaneously in an SU(2) effective chiral lagrangian. Whereas in HBChPT
one expands in powers of $p$ analogous to the meson sector,
in the ``small scale expansion'' one sets up a {\em phenomenological}
expansion in the small
scale $\epsilon$ denoting a soft momentum or the quark-mass or the
nucleon-delta mass splitting $\Delta = M_\Delta - M_N$. $\Delta$ is a
new dimensionful parameter in the theory which stays {\em finite} in the chiral
limit. Strictly speaking
the ``small scale expansion'' therefore has to be regarded as a
{\em phenomenological extension} of pure (HB)ChPT. Given these caveats it
should
be evident that the addition of spin-3/2 resonances as explicit degrees of
freedom to spin 1/2 HBChPT does not lead to one unique lagrangian as in
HBChPT, but is dependent on the expansion scheme one is employing. Possible
other expansion schemes that come to mind are the $SU(6)$ limit, where nucleon
and delta states are mass-degenerate to leading order
and the ``heavy resonance'' limit,
where $\Delta$ would count as a dimensionful parameter of order $p^0$.
For interesting and extensive work regarding the spin 3/2 resonances in
the large $N_c$ limit we refer the reader to ref.\cite{DJM94}
In this paper we focus exclusively on the ``small scale expansion''.

In the next section we present the successful ``heavy mass''
expansion method developed in refs. \cite{BKKM92,MRR92} for the simple
pedagogical example of
non-relativistic spinor electrodynamics and review spin 1/2 HBChPT.
In the following
section 3, we then show how this procedure can be generalized to deal with
spin 3/2. Section 4 gives the main
results of our formal development---giving a consistent chiral
perturbative scheme for the joint interactions of pions, nucleons, and
deltas to all orders.
In section 5, we review the construction of counterterm contributions in the
``small scale expansion'' and give the pertinent lagrangians to
${\cal O}(\epsilon^2)$.
We conclude our discussion at this point.
First applications of this formalism can be found
in \cite{HHK96a,HHK97,HHKK97,Menu97,chpt97appl}.

\section{1/M Expansion for Spin 1/2 Systems}

\subsection{Non-relativistic Spin 1/2 Electrodynamics}

We begin with a brief review of heavy mass techniques
for spin 1/2 systems. \cite{BKKM92,MRR92}  First, for simplicity, we
assume that only electromagnetic interactions are included.
Representing the nucleon as the two-component isospinor
\begin{equation}
\psi_N=\left(\begin{array}{c}
\psi_p\\
\psi_n
\end{array}\right)\nonumber
\end{equation}
with mass parameter $M_0$, the relativistic spin 1/2 Lagrangian has the
familiar form
\begin{equation}
{\cal L}_N=\bar{\psi}_N {}^{1\over 2}\Lambda^\gamma\psi_N \; .
\end{equation}
For a simple model of spinor electrodynamics where the nucleons do not posses
any anomalous magnetic moments one finds
\begin{equation}
{}^{1\over 2}\Lambda^\gamma=i\not\!\!{D}-M_0 \; , \label{eq:3}
\end{equation}
with the usual Dirac operator
\begin{equation}
D_\mu=\partial_\mu +i{e\over 2}(1+\tau_3)A_\mu
\end{equation}
being the covariant derivative. Here $e$ represents the
proton charge and is a positive quantity. In general, the four-momentum
$p_\mu$ can be written as
\begin{equation}
p_\mu=M_0v_\mu+k_\mu
\label{eq:a}
\end{equation}
where $v_\mu$ is a four-velocity satisfying $v^2=1$ and $k_\mu$ is a
soft momentum satisfying $k_\mu << M_0, \Lambda_\chi$ for all $\mu =0,1,2,3$.
By use of the operators
\begin{equation}
P_\pm={1\over 2}(1\pm\rlap/{v})
\end{equation}
we define the ``large''--$N$--and ``small''--$h$--components of the
nucleon field $\psi_N$ via the relations
\begin{eqnarray}
N(x)&\equiv& \exp(iM_0v\cdot x)P_+\psi_N(x) \; , \nonumber\\
h(x)&\equiv& \exp(iM_0v\cdot x)P_-\psi_N(x)\; ,
\end{eqnarray}
where we have included also the factor $\exp(iM_0v\cdot x)$ in order
to eliminate the mass dependence in the time development factor.

The nucleon Lagrangian then assumes the form
\begin{equation}
{\cal L}_N=\bar{N}{\cal A}_NN+\bar{h}{\cal B}_NN+\bar{N}\tilde{\cal
B}_Nh-\bar{h}{\cal C}_Nh  \; , \label{eq:nn}
\end{equation}
where
\begin{eqnarray}
{\cal A}_N&=&P_+({}^{1\over 2}\Lambda^\gamma +M_0\rlap/{v})P_+\; ,\nonumber\\
{\cal B}_N&=&P_-({}^{1\over 2}\Lambda^\gamma +M_0\rlap/{v})P_+\; ,\nonumber\\
\tilde{\cal B}_N&\equiv&\gamma_0{\cal B}_N^\dagger\gamma_0={\cal B}_N
\; , \nonumber\\
{\cal C}_N&=&-P_-({}^{1\over 2}\Lambda^\gamma +M_0\rlap/{v})P_- \; .
\end{eqnarray}
Using the projection operator identities
\begin{equation}
\begin{array}{cc}
P_\pm P_\mp=0\; , \quad P_\pm P_\pm=P_\pm \; , \quad
P_\pm\not\!\!{D}P_\pm=\pm v\cdot DP_\pm \; , \quad P_\pm\not\!\!{D}P_\mp
=\not\!\!{D}^\perp \; ,
\end{array}
\end{equation}
one obtains
\begin{eqnarray}
{\cal A}_N&=&iv\cdot D \; ,\nonumber\\
{\cal B}_N&=& i\not\!\!{D}^\perp=\tilde{\cal B}_N \; ,\nonumber\\
{\cal C}_N&=& 2M_0+iv\cdot D \; , \label{eq:Mgamma}
\end{eqnarray}
where
\begin{equation}
\not\!\!{D}^\perp=\not\!\!{D}-\rlap/{v}v\cdot D
\end{equation}
is the transverse component of $\not\!\!{D}$. From Eq.(\ref{eq:Mgamma}) one
can easily see that the original relativistic field $\psi_N$ has been
decomposed into a ``quasi-massless'' (``light'') field $N$ and a ``heavy''
field $h$ with a mass parameter of twice the nucleon mass. Quantizing via
path-integral methods, the functional integral
\begin{eqnarray}
W[{\rm sources}]&=&{\rm const.}\int[dN][d\bar{N}][dh][d\bar{h}]\exp i\int
d^4x\left( {\cal L}_N + {\rm source terms} \right)\nonumber\\
&&
\end{eqnarray}
can be diagonalized via the field-redefinition
\begin{equation}
h'\equiv h-{\cal C}_N^{-1}{\cal B}_NN \; ,
\end{equation}
{\it i.e.},
\begin{eqnarray}
W[{\rm sources}]&=&{\rm const.}\int[dN][d\bar{N}][dh'][d\bar{h}']
\exp i\int d^4x \nonumber\\
& & \times \left(\bar{N}({\cal A}_N+\tilde{\cal
B}_N{\cal C}_N^{-1}{\cal B}_N)N-\bar{h}'{\cal C}_Nh'+ {\rm source\
terms} \right) \nonumber \\
&=&{\rm const.}\int[dN][d\bar{N}] det\left({\cal C}_N\right)
\exp i\int d^4x \nonumber\\
& & \times \left(\bar{N}({\cal A}_N+\tilde{\cal
B}_N{\cal C}_N^{-1}{\cal B}_N)N+ {\rm source\ terms} \right) \; .
\end{eqnarray}
The determinant can be shown to yield a constant \cite{MRR92}, leaving
an effective lagrangian written only in terms of the light
components--$N$
\begin{equation}
{\cal L}_{\rm eff}=\bar{N}({\cal A}_N+\tilde{\cal B}_N
{\cal C}_N^{-1}{\cal B}_N)N \; .
\end{equation}
For our simple example of spin 1/2 electrodynamics the inverse operator
${\cal C}_N^{-1}$ can be expressed as a series
\begin{equation}
{\cal C}_N^{-1}={1\over 2M_0}\sum_{n=0}^\infty
\left({-iv\cdot D\over 2M_0}\right)^n \; , \label{eq:Cem}
\end{equation}
yielding the desired form for the effective action in terms of an
expansion in powers of $1/M_0$. The large nucleon mass
$M_0$ has been moved into interaction vertices,
thus providing a convenient theory with only ``light'' degrees of freedom.

At lowest order we have simply
\begin{equation}
{\cal L}_{\rm eff}^{(1)}=\bar{N}{\cal A}_NN=\bar{N}(iv\cdot\partial-{e\over
2}(1+\tau_3)v\cdot A)N \; , \label{eq:mm}
\end{equation}
resulting in the leading order (free) nucleon
propagator
\begin{equation}
D_N(k)={i\over v\cdot k+i\epsilon} \; , \label{eq:prop1}
\end{equation}
where $k$ is the soft momentum defined in Eq.(\ref{eq:a}).

The next higher order is then given by the first 1/M correction in
Eq.(\ref{eq:Cem}). One finds
\begin{equation}
{\cal L}_{\rm eff}^{(2)}=\bar{N}\tilde{\cal B}_N^{(1)}({\cal
C}_N^{-1})^{(0)}{\cal B}_N^{(1)}N=
\bar{N}{(i\not\!\!{D}^\perp)^2\over 2M_0}N \; .
\label{eq:b}
\end{equation}
Defining the Pauli-Lubanski spin vector
\begin{equation}
S_\mu={i\over 2}\gamma_5\sigma_{\mu\nu}v^\nu
\end{equation}
which obeys the following (d-dimensional) relations
\begin{eqnarray}
&&S\cdot v=0, \quad \{S_\mu, S_\nu\}={1\over 2}(v_\mu v_\nu - g_{\mu\nu}),
\quad S^2={(1-d)\over 4}, \nonumber \\
&&[ S_\mu, S_\nu ] = i\epsilon_{\mu\nu\alpha\beta}v^\alpha S^\beta\; ,
\quad (d=4)
\end{eqnarray}
Eq.(\ref{eq:b}) can then be written in the form
\begin{equation}
{\cal L}_{\rm eff}^{(2)}={1\over 2M_0}\bar{N}\left\{(v\cdot D)^2-D^2
+[ S_\mu ,S_\nu ][ D^\mu ,D^\nu ]\right\} N \; . \label{eq:em2}
\end{equation}
We recognize the spin-independent term then as a correction to the leading
order propagator structure Eq.(\ref{eq:prop1})
and its spin-dependent partner as the Dirac component of the
magnetic moment. If the spin 1/2 nucleons are to posses an additional
{\em anomalous} magnetic moment, the latter piece must be augmented by a
counterterm of the same form. This is discussed in sections 4, 5.
However, the spin-independent piece in Eq.(\ref{eq:em2}) must not be modified
by any counterterm contributions. This is
associated with ``reparametrization invariance'' which necessitates
that although there exists a freedom under the way in which
the momentum $p_\mu$ is decomposed into its four-velocity and
soft-momentum components, the square of the four-momentum must remain
invariant. \cite{LM92} That is to say, if one observer uses
$p_\mu=M_0v_\mu+k_\mu$ while another defines $p_\mu=M_0{v'}_\mu+{k'}_\mu$
it is required that
\begin{equation}
2M_0v\cdot k+k^2=2M_0v'\cdot k'+{k'}^2 \; . \label{eq:identity}
\end{equation}
Since the
leading piece of the effective action in momentum-space involves $v\cdot k$
we know that
it must be accompanied by a next order term $k^2\over 2M_0$ and that the
coefficient of this term must be unity for the identity
Eq.(\ref{eq:identity}) to hold.
A different way to convince oneself that the spin-independent terms in
Eq.(\ref{eq:em2}) are protected from extra counterterm contributions for a
wide class of theories has to do with the 2-photon (seagull) piece of
${\cal L}_{\rm eff}^{(2)}$ which is quadratic in
$eA_\mu$. This structure  must generate the familiar Thomson scattering
amplitude for a nucleon, whose form is required by rigorous low energy
theorems to be \cite{Thirring}
\begin{equation}
{\rm Amp}_{\gamma\gamma NN}={e^2\over M_0}
\epsilon\cdot\epsilon' \bar{N}{1\over 2}(1+\tau_3)N
\end{equation}
Higher order lagrangians ${\cal L}_{N}^{(n)}, \; n \geq 3$ in our example of
non-relativistic spin 1/2 electrodynamics can be obtained
via Eq.(\ref{eq:Cem}) and are suppressed by powers of $1/M_{0}^{(n-1)}$.
The infinite series recovers the full relativistic theory Eq.(\ref{eq:3}).

Having become familiar with the formalism of (non-relativistic) heavy
mass expansions in the simple example of spin 1/2 electrodynamics, it is now
straightforward to move on to the example of HBChPT.

\subsection{1/M Expansion in HBChPT}

Interactions with pions and with general external axial/vector fields
${\bf v_\mu}$,$v_\mu^{(s)}$,${\bf a_\mu}$ can be included in a
chiral-invariant
fashion via the operator \cite{GSS88}
\begin{equation}
{}^{1\over 2}\Lambda^{(1)}=i\not\!\!{D}-M_0+{g_A\over
2}\rlap/{u}\gamma_5\label{eq:qq}
\end{equation}
where
\begin{eqnarray}
U&=&u^2=\exp({i\over F_\pi}\vec{\tau}\cdot\vec{\phi})\nonumber \\
D_\mu N&=&(\partial_\mu+\Gamma_\mu -i v_{\mu}^{(s)})N\nonumber \\
\Gamma_\mu&=&{1\over 2}[u^\dagger,\partial_\mu u]-{i\over
2}u^\dagger({\bf v_\mu}+{\bf a_\mu})u-{i\over 2}u({\bf v_\mu}-{\bf
a_\mu})u^\dagger\equiv \tau^i \; \Gamma_{\mu}^i \nonumber \\
u_\mu&=&iu^\dagger\nabla_\mu Uu^\dagger\equiv \tau^i \; w_{\mu}^i
\nonumber \\
\nabla_\mu U&=&\partial_\mu U-i({\bf v_\mu}+{\bf a_\mu})U+iU({\bf v_\mu}-
{\bf a_\mu}).
\label{eq:qqq}
\end{eqnarray}
The effective action can be generated as before but now with
\begin{eqnarray}
{\cal A}^{(1)}_N&=&iv\cdot D+g_AS\cdot u\nonumber\\
{\cal B}^{(1)}_N&=&i\not\!\!{D}^\perp-{1\over 2}g_A(v\cdot u)\gamma_5
                    \nonumber\\
{\cal C}^{(0)}_N&=&2M_0 \nonumber \\
{\cal C}^{(1)}_N&=&iv\cdot D+g_AS\cdot u
\end{eqnarray}
To lowest---${\cal O}(p)$---order there exists a linear coupling to the
isovector axial field $\vec{a}_\mu$
\begin{equation}
{\rm Amp}_{NNa}=g_A\bar{N}S^\mu\vec{a}_\mu\cdot\frac{\vec{\tau}}{2} N
\end{equation}
and we recognize $g_A=1.26$ as the conventional nucleon
axial vector coupling constant.

At next---${\cal O}(p^2)$---order we find
\begin{eqnarray}
{\cal L}_{\rm eff}^{(2)}&=&\tilde{\cal B}_N^{(1)}
({\cal C}_N^{-1})^{(0)}{\cal B}_N^{(1)}\nonumber \\
&=&{1\over 2M_0}\bar{N}\left[(v\cdot D)^2-D^2
+[S_\mu,S_\nu][D^\mu,D^\nu]\right.\nonumber \\
& &\left.\phantom{{1\over 2M_0}\bar{N}}
-ig_A(S\cdot Dv\cdot u+v\cdot u S\cdot D)
-{1\over 4}g_A^2(v\cdot u)^2\right]N \label{eq:1mnn}
\end{eqnarray}
Here the meaning of the first three terms was given above for the case
of electromagnetic coupling. However, the $D_\mu$ terms now denote {\em chiral}
covariant derivatives containing new pion and axial vector source couplings in
addition to the (minimal) photon coupling discussed in the previous section.
The fourth
piece gives the Kroll-Ruderman term in charged pion
photoproduction\cite{KR54} while the last term starts out as a two-pion
coupling to the nucleon with the coupling strength fixed by $g_A$.

\section{1/M Expansion for Spin 3/2 Systems}

Before launching into the parallel discussion of the heavy baryon
treatment of spin 3/2 systems it is useful to first note a technical
but important point about the characterization of such systems.

\subsection{Point-invariance and Spin 3/2}

We begin with the standard relativistic form of the Lagrangian for a spin 3/2
field $\Psi_{\mu}$ and (bare) mass parameter $M_\Delta$.
Throughout this work spin 3/2 fields are represented via a
Rarita-Schwinger spinor \cite{RS}, for an overview of this formalism see
Appendix A. We have
\begin{equation}
{\cal L}_{3/2} = \bar{\Psi}^{\alpha}{}^{3\over 2}\Lambda(A)_{\alpha
\beta} \Psi^{\beta}
\label{eq:l3/2}
\end{equation}
where
\begin{eqnarray}
{}^{3\over
2}\Lambda(A)_{\alpha\beta}&=&-\left[(i\rlap/{\partial}-M)g_{\alpha\beta}
+iA(\gamma_\alpha\partial_\beta+\gamma_\beta\partial_\alpha)\right. \\
& &\phantom{-} \left. \; +{i\over
2}(3A^2+2A+1)\gamma_\alpha\rlap/{\partial}\gamma_\beta
+M_\Delta(3A^2+3A+1)\gamma_\alpha\gamma_\beta\right] \nonumber
\end{eqnarray}
contains a free and unphysical (``gauge'') parameter
$A$ $(A\neq-{1\over 2})$.
The origin of this parameter-dependence lies in the feature that such
relativistic spin 3/2 systems must be invariant under the so called
``point - transformation'' \cite{MC56,NEK71}
\begin{eqnarray}
\Psi_{\mu} ( x ) & \rightarrow & \Psi_{\mu} (x) + a \gamma_{\mu} \gamma_{\nu}
                                \Psi^{\nu} (x) \nonumber\\
A               & \rightarrow & \frac{ A - 2 a}{1 + 4 a} \label{eq:pti}
\end{eqnarray}
which simply says that an arbitrary admixture $a$ $(a\neq -1/4)$
of ``spurious'' or ``off-shell''
spin 1/2 components
$\gamma_\nu\Psi^\nu(x)$ which are {\em always} present in the relativistic
spin 3/2 field
$\Psi_{\mu} (x)$ can be compensated by a corresponding
change in the parameter $A$ to leave the Lagrangian
Eq.(\ref{eq:l3/2}) invariant. For $A=-1/3$ one recovers the original lagrangian
of Rarita and Schwinger \cite{RS}, whereas in the more recent literature
one tends to use $A=-1$, see {\it e.g.}
\cite{benmerrouche}. Physical quantities
are of course guaranteed to be independent of the choice of the $A$ parameter
by virtue of the ``KOS-theorem'' \cite{KOS}.

For our purposes it is convenient to employ a form for the relativistic
spin 3/2 Lagrangian that
was written down by Pascalutsa \cite{Pas94}. The advantage of this
formulation, when compared
with the more familiar one used in ref. \cite{benmerrouche} for
example, lies in
the feature that it allows the absorption of any dependence on the unphysical
parameter $A$ into a matrix $O_{\mu \nu}^{A}$, resulting in the form
\begin{equation}
{\cal L}_{3/2} = \bar{\Psi}^{\alpha} O_{\alpha \mu}^{A}
{}^{3\over 2}\Lambda^{\mu \nu}
             O_{\nu \beta}^{A} \Psi^{\beta}
\end{equation}
where
\begin{equation}
O_{\alpha \mu}^{A} =  g_{\alpha \mu} + \frac{1}{2} A \gamma_{\alpha}
                      \gamma_{\mu } \label{eq:moa}
\end{equation}
contains all the $A$-dependence leaving
\begin{eqnarray}
{}^{3\over 2}\Lambda_{\alpha\beta}^{free}&=&-[i(\rlap/{\partial}-M_\Delta)
      g_{\alpha\beta}
     -{1\over 4}\gamma_\alpha\gamma_\lambda(i\rlap/{\partial}-M_\Delta)
\gamma^\lambda\gamma_\beta]
\end{eqnarray}
as the $A$-independent leading-order free spin 3/2 lagrangian with the
redefined spin 3/2 field
\begin{equation}
\psi_{\mu} (x) = O^{A}_{\mu \nu} \Psi^{\nu} (x) \label{eq:physical} \; .
\end{equation}
$\psi_\mu$ is guaranteed to satisfy all point transformation requirements
by construction. Our
(non-relativistic) lagrangians written in terms of the $\psi_\mu$ fields
are therefore independent of $A$ and from now on we will
only work with the redefined fields $\psi_\mu$.

\subsection{Non-relativistic Spin 3/2 Electrodynamics}

Having addressed the requirement of point-invariance, it is useful to
begin the development of our formalism, as in the case of the
nucleon, by first dealing
with the inclusion only of minimal electromagnetic coupling.  In order
to do so, we must also take account of isotopic spin considerations.

\subsubsection{Isospin 3/2}

 Since
the $\Delta$ carries $I=3/2$ it is convenient to use a
(Rarita-Schwinger-like) isospurion notation, wherein an
isospin 3/2 state is described by an isospinor which carries also a
3-vector index and which is subject to the constraint
\begin{equation}
\tau^i \; \psi_{\mu}^i(x)=0\; , \quad i=1,2,3 \; ,
\end{equation}
where $\tau^i$ is a 2-component Pauli matrix in isospin space.
For details on this isospin formalism, the reader is referred to
Appendix B.  Thus the $\Delta$
field is described by the symbol $\psi_\mu^i(x)$, which is both a
Lorentz vector and spinor as well as an isotopic vector and
spinor.  In this case then
the relativistic Lagrangian describing the interaction of the delta
with the minimally coupled\footnote{We are not discussing the so-called
Rarita-Schwinger
gauges \cite{RS} here as they would imply constraints on the possible choices
for the
spin 1/2 admixture $A$. We will set up the formalism so that any choice of A
except for
$(A\neq -1/2)$ can be accommodated.} electromagnetic field is given by
\begin{equation}
{\cal L}_{S=3/2,I=3/2} = \bar{\psi}^{\mu}_{i} \;
^{3\over 2}\Lambda_{\mu \nu}^{(0)i j} \;
 \psi^{\nu}_{j} \label{eq:lagrange}
\end{equation}
where the operator
\begin{equation}
{}^{3\over
2}\Lambda_{\mu\nu}^{(0)ij}=-[(i\not\!\!{D}^{ij}-M_\Delta\delta^{ij})g_{\mu\nu}
-{1\over 4}\gamma_\mu
\gamma^\lambda(i\not\!\!{D}^{ij}-M_\Delta\delta^{ij})\gamma^\lambda\gamma_\nu]
\end{equation}
besides being 4x4 Dirac matrix and second rank Lorentz-tensor,
is also a 2x2 isotopic
matrix as well as a second rank tensor in isotopic spin space.  Here
and below Lorentz indices will always be denoted by Greek symbols
while isotopic spin indices will be designated by Roman letters.
The minimally coupled covariant derivative is given by
\begin{equation}
D_\mu^{ij}=\partial_\mu\delta^{ij}+ie[{1\over 2}(1+\tau_3)\delta^{ij}
-i\epsilon^{ij3}]A_\mu
\end{equation}
which is also a 2x2 matrix as well as a second-rank tensor in isotopic
spin space. The interested reader can easily verify that the
operator
\begin{equation}
I_3^{ij}={1\over 2}\tau_3\delta^{ij}-i\epsilon^{ij3}
\end{equation}
when acting on the state $\Psi^j$ given in Eq.(\ref{eq:rr})
merely multiplies each
component of the delta by its appropriate value of $I_3$---{\it e.g.}
\begin{equation}
I_3^{3i}\Psi^i=-\sqrt{2\over 3}\left[ \begin{array}{c}
{1\over 2}\Delta^+\\
-{1\over 2}\Delta^0
\end{array}\right]
\end{equation}
Having set up our Lagrangian we can proceed to the development of the
appropriate heavy baryon formalism.

\subsubsection{Transition to Heavy Baryon Fields}

In this section we discuss the transformation of the relativistic theory
of spin 3/2 particles into the corresponding heavy baryon form.  The
calculation is analogous to that given for the nucleon sector and
described in section 2.  However, it is also considerably more complex,
since the relativistic Lagrangian
formulation of spin 3/2 fields Eq.(\ref{eq:lagrange}) contains on-shell and
off-shell components, spin 3/2 and two independent
spin 1/2 degrees of freedom, large and small
Dirac spinor pieces, leading order and $1/M_{\Delta}$
contributions, etc., all combined in a very compact notation.
Our goal, as stated in the introduction,
is to disentangle all these contributions and to develop a systematic
$1/M_{\Delta}$ expansion in terms of a Lagrangian involving the
``large'' component spin 3/2-isospin 3/2 fields as the only explicit
degree of freedom, {\em without} losing any of
the physics contained in Eq.(\ref{eq:lagrange}).

Our first step is to identify the spin 3/2 and (two) spin 1/2 degrees of
freedom.  Using the (appropriately modified) spin projection operators
from the
relativistic theory---{\it cf.} Eq.(\ref{eq:zz})---we introduce a complete
set of orthonormal spin projection operators\footnote{In the theory of
heavy quarks one also has to deal with the problem of spin 3/2 baryons.
For a discussion of the use of projection operators in HQET we refer
to the article by Falk \cite{Fal92}.}
\begin{eqnarray}
_{(33)}P^{3/2}_{\mu \nu} & = & g_{\mu \nu} - \frac{1}{3} \gamma_{\mu} \gamma_{
                       \nu} - \frac{1}{3} \left( \not\!{v} \gamma_{\mu}
                       v_{\nu} + v_{\mu} \gamma_{\nu} \not\!{v} \right)
                       \nonumber\\
_{(11)}P^{1/2}_{\mu \nu} & = & \frac{1}{3} \gamma_{\mu} \gamma_{\nu} - v_{\mu}
                               v_{\nu} + \frac{1}{3} \left( \not\!{v}
                               \gamma_{\mu} v_{\nu} + v_{\mu} \gamma_{\nu}
                               \not\!{v} \right)
                               \nonumber\\
_{(22)}P^{1/2}_{\mu \nu} & = & v_{\mu} v_{\nu} \nonumber\\
_{(12)}P^{1/2}_{\mu \nu} & = & \frac{1}{\sqrt{3}} \left( v_{\mu} v_{\nu} -
                               \not\!{v} v_{\nu} \gamma_{\mu} \right)
                               \nonumber\\
_{(21)}P^{1/2}_{\mu \nu} & = & \frac{1}{\sqrt{3}} \left( \not\!{v} v_{\mu}
                               \gamma_{\nu} - v_{\mu} v_{\nu} \right)
                               \label{eq:yy}
\end{eqnarray}
obeying
\begin{eqnarray}
_{(33)}P^{3/2}_{\mu \nu} + _{(11)}P^{1/2}_{\mu \nu} &+& _{(22)}P^{1/2}_{\mu
\nu}
                        =  g_{\mu \nu}  \nonumber\\
_{(ij)}P^{I}_{\mu \nu} \; _{(kl)}{P^{J, \nu}}_\delta
                       & = & \delta^{I J} \; \delta_{jk}
\; _{(il)}P^{J}_{\mu\delta}
\end{eqnarray}
With the use of these projectors we
split up $\psi^{i}_{\mu}$ into three independent degrees of freedom
\begin{eqnarray}
^{(1)}\psi^{i}_{\mu} (x) & = & _{(11)}P^{1/2}_{\mu \nu}
                               \psi^{\nu}_{j} (x) \nonumber\\
^{(2)}\psi^{i}_{\mu} (x) & = & _{(22)}P^{1/2}_{\mu \nu}
                               \psi^{\nu}_{j} (x) \nonumber\\
^{(3)}\psi^{i}_{\mu} (x) & = & _{(33)}P^{3/2}_{\mu \nu}
                               \psi^{\nu}_{j} (x)
\end{eqnarray}
in terms of which the operator ${}^{3\over 2}\Lambda^{(0)}_{\mu\nu}$, which
defines the
relativistic Lagrangian, becomes a 3x3 matrix, with
\begin{eqnarray}
{}^{3\over 2}_{(33)}\Lambda^{(0)\mu \nu}
& = & - \left[ ( i \not\!\!{D} - M_{\Delta} ) g^{\mu
                               \nu} \right ] \nonumber\\
{}^{3\over 2}_{(31)}\Lambda^{(0)\mu \nu}
& = & {}^{3\over 2}_{(13)}\Lambda^{(0)\mu \nu} =
- \left[ i \not\!\!{D} g^{\mu \nu} \right] \nonumber\\
{}^{3\over 2}_{(23)}\Lambda^{(0)\mu \nu} & = & {}^{3\over
2}_{(32)}\Lambda^{(0)\nu \mu}=0 \nonumber\\
{}^{3\over 2}_{(11)}\Lambda^{(0)\mu \nu} & = & - \left[ ( i \not\!\!{D} -
M_{\Delta} ) g^{\mu
                               \nu} + \frac{i}{2} \gamma^{\mu} \not\!\!{D}
                               \gamma^{\nu} + 3M_\Delta g^{\mu\nu}
                               \right] \nonumber\\
{}^{3\over 2}_{(12)}\Lambda^{(0)\mu \nu} & = & - \left[ \frac{i}{2}
\gamma^{\mu} \not\!\!{D}
                               \gamma^{\nu} + \sqrt{3} M_{\Delta} \;
                               _{(12)}P^{\mu \nu}_{1/2} \right] \nonumber\\
{}^{3\over 2}_{(21)}\Lambda^{(0)\mu \nu} & = & - \left[ \frac{i}{2}
\gamma^{\mu} \not\!\!{D}
                               \gamma^{\nu} + \sqrt{3} M_{\Delta} \;
                               _{(21)}P^{\mu \nu}_{1/2} \right] \nonumber\\
{}^{3\over 2}_{(22)}\Lambda^{(0)\mu \nu} & = & - \left[ ( i \not\!\!{D} -
M_{\Delta} ) g^{\mu
                               \nu} + \frac{i}{2} \gamma^{\mu} \not\!\!{D}
                               \gamma^{\nu} + M_{\Delta} g^{\mu \nu}
                               \right] \label{eq:22Lambda}
\end{eqnarray}

The leading order contribution here is
${}^{3\over 2}_{(33)}\Lambda^{(0)\mu \nu}$, as this is the only piece which
involves
only spin 3/2 degrees of freedom.   However, it
remains to decompose this field into large and small components and to
eliminate the $M_\Delta$ dependence in the time development factor as
in the analogous nucleon calculation.  Thus we define ``large''
and  ``small'' spin 3/2 fields via
\begin{eqnarray}
T^{i}_{\mu} (x)                & = &  P_{+} \; ^{(3)}\psi_{\mu}^{i} (x)
                                      \; \mbox{exp}(i M_\Delta v \cdot x)
                                      \nonumber\\
^{(3-)}h^{i}_{\mu} (x)         & = &  P_{-} \; ^{(3)}\psi_{\mu}^{i} (x)
                                      \; \mbox{exp}(i M_\Delta v \cdot
x) \label{eq:T}
\end{eqnarray}
which satisfy the constraints
\begin{eqnarray}
v_{\mu} T^{\mu}_{i} = & \gamma_{\mu} T^{\mu}_{i} & = 0 \nonumber\\
v_{\mu} \;\; ^{(3-)}h^{\mu}_{i} = & \gamma_{\mu} \;\;
^{(3-)}h^{\mu}_{i} & = 0 \label{eq:xx}
\end{eqnarray}
Eq.(\ref{eq:xx}) gives the heavy baryon analogues to the subsidiary
condition Eq.(\ref{eq:subsidiary}) for the relativistic
Rarita-Schwinger spinor of Appendix A. However, we note that
$\partial_{\mu} T^{\mu}_{i} = 0$ is in general not true in the heavy baryon
formalism, unlike in the relativistic case Eq.(\ref{eq:subsidiary2})

Here the fields $T^{\mu}_{i}(x)$ are the (field-redefined) SU(2) isospin
analogues of the
decuplet fields used in ref.\cite{JM91b}, whereas the $^{(3-)}h^{\mu}_i$
degrees of freedom are not considered there.
In addition,
if we are not content with just the leading order heavy baryon spin
3/2 Lagrangian, then we must also introduce heavy baryon fields
for the (off-shell) spin 1/2 components via
\begin{equation}
^{(\alpha \pm)}h^{i}_{\mu} (x) = P_{\pm} \;\; ^{(\alpha)}\psi_{\mu}^{i} (x)
                                 \; \mbox{exp}(i M_\Delta v \cdot x),
\label{eq:alphah}
\end{equation}
with $\alpha = 1,2$ labeling the independent spin-1/2 sectors with which
we are dealing.

The six resulting degrees of freedom for a spin 3/2 heavy baryon formalism are
therefore $T_{i}^{\mu} (x)$ and
\begin{equation}
G^{i}_{\mu} = \left(\begin{array}{c}
 ^{(3-)}h^{i}_{\mu}\\
 ^{(1+)}h^{i}_{\mu}\\
^{(2-)}h^{i}_{\mu}\\
^{(1-)}h^{i}_{\mu}\\
^{(2+)}h^{i}_{\mu}
\end{array}\right) , \label{eq:Gmu}
\end{equation}
where the small component spin 3/2 field and the spin 1/2 contributions
have for notational convenience been combined in the column vector
$G_{i}^{\mu} (x)$.  Suppressing all four-vector
and isospin indices the heavy baryon Lagrangian corresponding to
Eq.(\ref{eq:lagrange}) then can be written in the simple form
\begin{equation}
{\cal L}_{\Delta} = \bar{T} {\cal A}_{\Delta} T + \bar{G}
             {\cal B}_{\Delta} T + \bar{T} \gamma_{0}
             {\cal B}_{\Delta}^{\dagger} \gamma_{0} G - \bar{G}
             {\cal C}_{\Delta} G, \label{eq:4terms}
\end{equation}
where ${\cal A}_{\Delta}, {\cal B}_{\Delta},
{\cal C}_{\Delta}$ are all matrices containing the (electromagnetic) covariant
derivative in our example of (non-relativistic) spin 3/2 electrodynamics
and whose explicit forms will be given in the next section.  However,
while Eq.(\ref{eq:4terms}) looks very much like its nucleonic analogue
Eq.(\ref{eq:nn}), it
must be kept in mind that there are also important differences.
Specifically, in terms of
our six-component decomposition into large and small spin 3/2 and 1/2
fields ${\cal A}_\Delta$ is a number, ${\cal B}_\Delta$ is
a five-component column vector, while ${\cal C}_\Delta$ is a
5x5 matrix.

\subsubsection{The Interaction with Photons}

We can now proceed to determine the desired effective Lagrangian in terms
of the ``large'' spin 3/2 field $T^\mu$ by integrating out the $G^\mu$ degrees
of freedom just as done for the analogous nucleonic system---shifting
$G^\mu$ fields via
\begin{equation}
G \rightarrow G' + {\cal C}_{\Delta}^{-1} \; {\cal B}_{\Delta}^{(1)}
\; T \label{eq:Gprime2}
\end{equation}
the heavy baryon Lagrangian is diagonalized. We then integrate out the
unwanted $G'$ fields, yielding
\begin{equation}
{\cal L}_{\Delta} = \bar{T}\left[ {\cal A}_\Delta +\gamma_{0}
{\cal B}_{\Delta}^\dagger \gamma_{0} {\cal C}_{\Delta}^{-1}
                       {\cal B}_{\Delta}\right] T \label{eq:p2}
\end{equation}
which is the result we seek. Given that we are only interested in the
next-to-leading order corrections to the ${\cal O}(\epsilon)$
lagrangians in this paper the process of ``integrating out'' the unwanted
fields is trivial. It just amounts to dropping the (decoupled) $G^\prime$ terms
in the lagrangian. However, in general this procedure can be more involved.
To carry this procedure to ${\cal O}(\epsilon^3)$ one carefully has to keep
all the source dependencies to obtain the proper wavefunction renormalization
({\it e.g. \cite{EM97}}). Furthermore, the resulting determinant in general
need not be equal to unity anymore as in standard HBChPT because it contains
particle and anti-particle excitations of the spin 1/2 (off-shell) components
of the relativistic delta theory. A detailed study of these and other effects
is in preparation \cite{HHK97c} but the ${\cal O}(\epsilon^2)$ results given
in this paper should remain unaffected.

Specifically, the leading order (non-relativistic)
$\gamma\Delta\Delta$
lagrangian is simply given
by the term
\begin{equation}
{\cal L}_{\Delta\Delta\gamma}^{(1)} = \bar{T}^{\mu}_{i} (x) \;
                         {\cal A}_{\Delta , \mu \nu}^{ij} \;
                         T^{\nu}_{j} (x) \label{eq:addg}
\end{equation}
where
\begin{equation}
{\cal A}_{\Delta,\mu\nu}^{ij}=-i\; v\cdot D^{ij}\; g_{\mu\nu} \; .
\end{equation}
The (free)
leading order $\Delta$  propagator in the spin 3/2 isospin 3/2 subspace
is then
\begin{equation}
D_F(p)_{\mu\nu}^{ij}={-i\over v\cdot
k+i\epsilon}\; P_{\mu\nu}^{3/2} \; \xi^{ij}_{3/2} \; , \label{eq:p3}
\end{equation}
with the d-dimensional non-relativistic spin 3/2 projector
\begin{eqnarray}
P_{\mu\nu}^{3/2}&\equiv &P_{v}^+ \; _{(33)}P_{\mu\nu}^{3/2} \; P_{v}^+
                         \nonumber \\
                &=&g_{\mu\nu}-v_\mu v_\nu + \frac{4}{d-1}S_\mu S_\nu
\end{eqnarray}
and the isospin 3/2 projector $\xi^{ij}_{3/2}$ defined in Appendix B.
Note that we have split up the $\Delta$ momentum as
\begin{equation}
p_\mu=M_\Delta v_\mu+k_\mu\; ,
\end{equation}
where $k_\mu$ refers to off-shell momentum in the propagator Eq.(\ref{eq:p3}).

The first higher order---$1/M_\Delta$---correction then has the form
\begin{equation}
{\cal L}_{\Delta\Delta\gamma}^{(2)}=\bar{T}\tilde{\cal B}_\Delta({\cal
C}_\Delta^{-1})^{(0)}{\cal B}_\Delta T
\end{equation}
which looks very much like the corresponding nucleon expression
Eq.(\ref{eq:mm})
except that ${\cal B},{\cal C}$ are vector, matrix quantities as
emphasized above.  The column vector ${\cal B}$ is straightforwardly
found to be
\begin{equation}
{\cal B}_{\Delta ,\mu\nu}^{ i j} = - \left(
               \begin{array}{c}
                P_{-} \left[ _{(33)}P^{3/2}_{\mu \nu} i \not\!\!{D}^{ij}_{
                          \perp} \right] P_{+} \\ \\
                 0  \\ \\
                 0    \\ \\
                P_{-} \left[ _{(11)}P^{1/2}_{\mu \nu} i \not\!\!{D}^{ij}_{
                          \perp} \right]  P_{+} \\ \\
                          0
               \end{array}\right)
\label{eq:BDelta}
\end{equation}
However, determining the inverse of the 5x5 matrix ${\cal C}_\Delta$ presents
more of a challenge.  Following our procedure in the case of the
nucleon, we shall use a representation in terms of a
power series in $(1/M_{\Delta})^n$.  Thus in order to find the leading
piece $(C_\Delta^{-1})^{(0)}$
we need only take
into account the terms in matrix ${\cal C}_{\Delta}$ that
depend
explicitly on
$M_{\Delta}$.  These mass dependent terms in
${\cal C}_{\Delta}$ are schematically displayed in Table 1, with
$"."$ denoting kinetic energy contributions. One
recognizes a convenient block-diagonal structure in the mass terms, which
facilitates the construction of the inverse matrix.   The explicit
construction for
${\cal C}_{\Delta}^{-1}$ at ${\cal O
}(1/M_{\Delta}$) is given in Appendix C, refuting claims of its non-existence
\cite{Belkov} by explicit construction.

\begin{table}
\hspace{4cm}
\begin{tabular}{r|ccccc}
   & 3-  & 1+  & 2-  & 1-  & 2+  \\ \hline
3- & M   & .   & .   & .   & .   \\
1+ & .   & M   & M   & .   & .   \\
2- & .   & M   & M   & .   & .   \\
1- & .   & .   & .   & M   & M   \\
2+ & .   & .   & .   & M   & M
\end{tabular}
\caption{$M_\Delta$ dependent terms in the matrix ${\cal C}_{\Delta}$}
\end{table}

Putting all these results together we arrive finally at the
first non-relativistic correction to the leading order
minimally coupled $\gamma\Delta\Delta$ Lagrangian Eq.(\ref{eq:addg})
\begin{eqnarray}
{\cal L}_{\Delta\Delta\gamma}^{(2)}
&=&-{1\over 2M_\Delta}\bar{T}^\mu_i(x)[(v\cdot D^{ik}
v\cdot D^{kj}-
D_\alpha^{ik}D_\beta^{kj}g^{\alpha\beta} )g_{\mu\nu}\nonumber\\
&+&
[S^\alpha,S^\beta](D_\alpha^{ik}D_\beta^{kj}-D_\beta^{ik}D_\alpha^{kj})
g_{\mu\nu}]T^\nu_j
\end{eqnarray}
The physics here is identical to the case of the nucleon---we
recognize immediately the
kinetic energy term and the interaction of the Dirac moment with
the electromagnetic field.  Likewise, from the former we can read off
the seagull term which generates the Thomson scattering amplitude
\begin{equation}
{\rm Amp}_{\gamma\gamma TT}=-{e^2\over M_\Delta}\epsilon\cdot\epsilon'\bar{
T}^i_\mu(\delta_{ij}{1\over 2}(3+\tau_3)
-i\epsilon_{3ij}(1+\tau_3)-\delta_{i3}\delta_{j3})g_{\mu\nu}T^j_\nu
\end{equation}
whose form is required by the low energy theorems.

\subsection{1/M Expansion for Chiral Spin 3/2 Lagrangians}

In the previous subsection we carried out the calculation in
some detail since when interactions with pions have to be included
the form becomes considerably more
complex and it is useful to have become familiar with the formalism in the
simpler minimally-coupled case.  For the leading order relativistic chiral
spin 3/2 lagrangian we employ the
matrix
\begin{eqnarray}
{}^{3\over 2}\Lambda_{\mu \nu}^{\pi i j} & = & - [ ( i \not\!\!{D}^{ij} -
                              M_{\Delta}\delta^{ij} ) g_{\mu \nu} -
                              \frac{1}{4}
                              \gamma_{\mu} \gamma^{\lambda} ( i
                              \not\!\!{D}^{ij} -
                              M_\Delta
                              \delta^{ij} ) \gamma_{\lambda} \gamma_{\nu}
                              \nonumber \\
                        &   & + \frac{g_{1}}{2} g_{\mu \nu} \not\!{u}^{ij}
                              \gamma_{5} + \frac{g_{2}}{2} ( \gamma_{\mu}
                              u_{\nu}^{ij} + u_{\mu}^{ij}
                              \gamma_{\nu} ) \gamma_{5}
                              \nonumber \\
                        &   & + \frac{g_{3}}{2} \gamma_{\mu} \not\!{u}^{ij}
                              \gamma_{5} \gamma_{\nu} ]  \label{eq:Lambda}
\end{eqnarray}
where
\begin{equation}
u^{ij}_\mu=\xi^{ik}_{3/2} \; u_\mu \; \xi_{3/2}^{kj}
\end{equation}
is used to generate the pion couplings\footnote{In the definition of
$u_\mu^{ij}$ one can have additional terms proportional to
$i\epsilon^{ijk}w_\mu^k$. However, one can absorb their contributions into
redefinitions of the coupling constants $g_i,\;i=1,2,3$ with the help of the
isospin algebra given in Appendix B.} . Note that
$D_\mu^{ij}$ is now the chiral covariant derivative which includes coupling to
the pions as well as to external vector,axial fields ${\bf v_\mu},
v_{\mu}^{(s)},{\bf a_\mu}$ via
\begin{equation}
D_\mu^{ij}=\partial_\mu\delta^{ij}+C_\mu^{ij}
\end{equation}
where
\begin{equation}
C^{ij}_\mu=\delta^{ij}\left(\Gamma_\mu-i v^{(s)}_\mu\right)
-2i\epsilon^{ijk}\Gamma^k_\mu
\end{equation}
We also have
extended the lagrangian of the previous section by appending
three interaction terms
with coupling constants $g_{1}, g_{2}, g_{3}$, which represent the most
general, chiral- and Lorentz-invariant way to couple a pion to a spin 3/2
particle.  Note that $g_{2},g_{3}$
only contribute if one,both of the spin 3/2 fields is,are
off-mass-shell. In the following it will become clear that off-shell parameters
such as $g_2,g_3$ do not pose a problem for the ``small scale expansion'' but
can
be treated on the same footing as any other coupling constant when performing
the
transition from the fully relativistic lagrangians to the non-relativistic
ones.
Although little if anything is known at this point about the
off-shell couplings $g_2,g_3$, one could in principle utilize the
SU(6) quark model in order to estimate the size of the physical pion-delta
coupling constant $g_{1}$, yielding
\begin{equation}
g_{1}= {9\over 5}g_{A} \; .
\end{equation}
However, we will treat $g_{1}$ on the same footing as any
other low energy constant in chiral perturbation theory and trust that in
subsequent work $g_1,g_2,g_3$ can be extracted from fits to experimental data.

It is then straightforward to use projection
operators to find the forms for the quantities
${\cal A}_\Delta,{\cal B}_\Delta,{\cal C}_\Delta$
\begin{eqnarray}
{\cal A}_{\Delta,\mu\nu}^{(1),ij}&=&
-[iv\cdot D^{ij}+g_1S\cdot u^{ij}]g_{\mu\nu} \label{eq:AB} \\
& & \nonumber \\
{\cal B}_{\Delta,\mu\nu}^{(1)ij}&=&-\left( \begin{array}{c}
P_-[{}_{(33)}P^{3/2}_{\mu\nu}i\not\!\!{D}^{\perp ij}-{1\over 2}g_1{}_{(33)}
P^{3/2}_{\mu\nu}v\cdot u^{ij}\gamma_5]P_+ \\
P_+[({2\over 3}g_1+g_2)S_\mu u^{ij}_\nu]P_+\\
P_-[-{1\over 2}g_2v_\mu u^{ij}_\nu\gamma_5]P_+\\
P_-[{}_{(11)}P^{1/2}_{\mu\nu}i\not\!\!{D}^{\perp ij}]P_+\\
0
\end{array}\right)
\end{eqnarray}
In the case of the matrix ${\cal C}^{-1}_\Delta$ we require only the
leading ${\cal O}(\epsilon^0)$ form already given in Appendix C since we are
working only to ${\cal O}(\epsilon^2)$.

To leading order we have then
\begin{equation}
{\cal L}_{\Delta\Delta}^{(1)} = \bar{T}^{\mu}_{i} (x) \;
                         {\cal A}_{\Delta , \mu \nu}^{{(1)} ij} \;
                         T^{\nu}_{j} (x)
\label{eq:ADelta}
\end{equation}
with ${\cal A}_{\Delta , \mu \nu}^{{(1)} ij}$ given in Eq.(\ref{eq:AB}).
The form of this leading order lagrangian should not come as a surprise.
For the special case of $v^\mu=(1,0,0,0)$ it corresponds to the
non-relativistic delta isobar model from the 1970s \cite{EW}. For a general
$v^\mu$ it reproduces the lagrangian of Jenkins and Manohar \cite{JM91b}.
All mass dependence is gone\footnote{Note that for the case of a coupled
spin 1/2 - spin 3/2 system in general one is left with a residual mass
dependence in the non-relativistic lagrangians. This will be discussed
in section 4.}, as desired. The
first term contains the kinetic energy of the spin 3/2 particle and
provides, among other things, a minimally coupled interaction
with the electromagnetic
field, whereas the second term starts out as a one pion vertex with derivative
coupling to the spin 3/2 fields.

At next---${\cal O}(\epsilon^2)$---order we find the
$1/M_{\Delta}$ Lagrangian for spin 3/2 particles as
\begin{eqnarray}
{\cal L}_{\Delta}^{(2)} & = &  \tilde{\cal B}_\Delta^{(1)}({\cal
C}_\Delta^{-1})^{(0)}{\cal B}_\Delta^{(1)}\nonumber\\
                &=&\frac{1}{2 M_{\Delta}} \; \bar{T}^{\mu}_{i} (x)
                           \left\{ \left[ \; D_{\alpha}^{ik}
                           D_{\beta}^{kj} \; g^{\alpha\beta} - v \cdot
                           D^{ik}  v \cdot D^{kj} \right] \; g_{\mu \nu}
\right.
                           \nonumber \\
                & &\phantom{\frac{1}{2 M_{\Delta}} \; \bar{T}^{\mu}_{i} (x)}
                           + g_{1} \; i \left( S \cdot D^{ik}  v
                           \cdot u^{kj} + v \cdot u^{ik}  S \cdot D^{kj}
                           \right) \; g_{\mu \nu}
                           \nonumber \\
                & &\phantom{\frac{1}{2 M_{\Delta}} \; \bar{T}^{\mu}_{i} (x)}
                           - \left[ S^{\alpha} , S^{\beta} \right] \left(
                           D_{\alpha}^{ik} D_{\beta}^{kj} - D_{\beta}^{ik}
                            D^{\alpha}_{kj} \right) \; g_{\mu\nu}
                           \nonumber \\
                & &\phantom{\frac{1}{2 M_{\Delta}} \; \bar{T}^{\mu}_{i} (x)}
                           + \frac{g_{1}^{2}}{4} v \cdot u^{ik}
                           v \cdot u_{kj} \; g_{\mu \nu}
                           \nonumber \\
                & &\phantom{\frac{1}{2 M_{\Delta}} \; \bar{T}^{\mu}_{i} (x)}
                           \left. - u_{\mu}^{ik} u_{\nu}^{kj} \;
                           \frac{g_{1}^{2} + 4 g_{1} g_{2} + 3 g_{2}^{2}}{3}
                           \right\} T^{\nu}_{j} (x) \; .
                           \label{eq:ss}
\end{eqnarray}
We can now discuss the physics contained in this new $1/M_{\Delta}$
Lagrangian for
spin 3/2 particles, only mentioning the most prominent features.
As before, the first term in Eq.(\ref{eq:ss}) contains the kinetic
energy and the two photon vertex which accounts for Thomson
scattering.  The second term corresponds to the Kroll-Ruderman term
in the nucleon sector,
i.e. it contains a one photon one pion vertex. Note that in our construction of
the ${\cal O}(\epsilon^2)$
counterterms in section 5 we do not expect to find a term
that has the same structure as either one of these two
expressions---neither vertex gets renormalized at ${\cal O}(\epsilon^{2})$
due to reparameterization
invariance \cite{LM92}, just as in the nucleon system. The third expression
gives a one photon vertex at ${\cal O}(1/M_{\Delta})$, which corresponds to a
photon coupled to the Dirac moment of a spin 3/2 particle. This vertex will
get renormalized by counterterms corresponding to anomalous magnetic
moments of spin 3/2 particles. Finally, the fourth and fifth term, among
other contributions, generate two-pion vertices at lowest order.
Note that the fifth
term does not have an analogue in the nucleon system and depends on the unknown
${\cal O}(\epsilon)$
off-shell coupling constant $g_{2}$ of Eq.(\ref{eq:Lambda}).

At this point we have achieved our goal of constructing the next to leading
order spin 3/2 Lagrangian for a pure spin 3/2 system. In principle one can
generalize our procedure to obtain even higher order corrections. However,
in most practical applications of this formalism one has to deal with the
somewhat more complex situation of having both spin 3/2 ($\Delta
(1232)$)
and spin 1/2 degrees of freedom (nucleons) present simultaneously.
We now shift the discussion to this more general case then,
noting that we will reutilize many of the structures that we
defined in this section.

\section{Simultaneous ${\bf 1/M}$ Expansion for Coupled Delta and
Nucleon Systems}

\subsection{Nucleon-Delta Transition Lagrangians}

For the non-relativistic reduction of the $\pi N\Delta$ system we first require
the relativistic lagrangians
which couple $N$ and $\Delta$ degrees of freedom and satisfy
invariance under ``point-transformations'' as discussed in section 3.1. We note
that
the general relativistic ${\cal O}(\epsilon^n)$ $N\Delta$ transition lagrangian
can be
written as
\begin{eqnarray}
{\cal L}_{\pi N\Delta}^{(n)}[A]&=&\bar{\psi}_N\; O_{\mu}^{(n)} \;
\Theta^{\mu\nu}_A
                                  \Psi_\nu + h.c. \; ,
\end{eqnarray}
where $O_{\mu}^{(n)}$ is a general chiral transition matrix to order
$\epsilon^n$
and $\Theta_{\mu\nu}^A$ is the most general (Dirac-) tensor that guarantees
``point-transformation'' invariance (Eq.(\ref{eq:pti}))
for the relativistic spin 3/2 Rarita-Schwinger field $\Psi_\mu$. Nath, Etemadi
and Kimel
\cite{NEK71} have determined this most general tensor to be
\begin{equation}
\Theta_{\mu\nu}^A=g_{\mu\nu}+\left[z_i+\frac{1}{2}\left(1+4z_i\right)A\right]\gamma_\mu
                  \gamma_\nu \; ,
\end{equation}
where $z_i$ denotes a free parameter which governs the coupling of the
(off-shell) spin 1/2 components to a given matrix $O_{\mu, \; i}^{(n)}$. Guided
by
the note by Pascalutsa \cite{Pas94}, we again observe that the $A$-dependence
factors out and write
\begin{eqnarray}
\Theta_{\mu\nu}^A&=&\left(g_{\mu\alpha}+z_i\gamma_\mu\gamma_\alpha\right)
                    \left(g^{\alpha}_\nu+\frac{A}{2}\gamma^\alpha\gamma_\nu\right)
                    \nonumber \\
                 &=&\Theta_{\mu\alpha}(z_i) \; O^{\alpha \; A}_\nu \; ,
\end{eqnarray}
with $O_{\alpha\beta}^A$ given by Eq.(\ref{eq:moa}). Therefore we can subsume
all
dependencies on the (unphysical) parameter $A$ via the same field redefinition
Eq.(\ref{eq:physical}) already introduced in our discussion of the pure spin
3/2
sector in section 3.1 and find
\begin{eqnarray}
{\cal L}_{\pi N\Delta}^{(n)}&=&\bar{\psi}_N \; O_{\mu, \; i}^{(n)} \;
                               \Theta^{\mu\nu}(z_i)
                               \; \psi_\nu + h.c. \; ,
\end{eqnarray}
where $\psi_\nu$ denotes the field-redefined spin 3/2 field of
Eq.(\ref{eq:physical}).

For the leading---${\cal O}(\epsilon)$--order relativistic $N\Delta$ transition
lagrangian we therefore write\footnote{We have made the assumption that the
leading
order $N\Delta$ lagrangian can be written down without terms that involve the
nucleon
or delta equations of motion. See section 5.3 for a more detailed discussion of
this
issue.}
\begin{equation}
                {\cal L}_{\pi N\Delta}^{(1)}  = g_{\pi N\Delta}
                                \left\{ \bar{\psi}^{\mu}_{i}
                                  \; \Theta_{\mu\alpha}(z_0) \; w^{\alpha}_{i}
\;
                                  \psi_N + \bar{\psi}_N \;
                                  w^{\alpha\dagger}_{i} \;
                                  \Theta_{\alpha\mu}(z_0) \; \psi^{\mu}_{i}
                                  \right\}
                                  \label{eq:NDelta}
\end{equation}
with
\begin{eqnarray}
\Theta_{\mu\nu}(z_0) & = & g_{\mu\nu} + z_0 \gamma_{\mu} \gamma_{\nu} \; .
\label{eq:Theta}
\end{eqnarray}
We therefore note that in the leading order {\em relativistic} $N\Delta$
lagrangian
one finds two independent $\pi N\Delta$ couplings---the so-called ``on-shell''
coupling constant $g_{\pi N\Delta}$ and the so-called ``off-shell parameter''
$z_0$.
While a determination of $g_{\pi N\Delta}$ within the ``small scale expansion''
is given
in the next section,
there is no agreement in the literature about the magnitude and status of
$z_0$.
Many years ago Peccei \cite{Peccei} argued for $z_0=-1/4$, while Nath, Etemadi
and Kimel
\cite{NEK71} determined $z_0=+1/2$ from requirements of local causality. More
recently
the RPI-group \cite{benmerrouche} has argued to leave all ``off-shell
parameters'' as
free parameters to be fitted from physical observables. In this work we will
adopt the
later approach and treat $z_0$ as a free parameter to be determined in future
applications of this theory, as our formalism does not {\em require} a specific
choice
for it.

\subsection{Leading Order Lagrangians}

The full relativistic Lagrangian that we must consider for a system of
nucleon, $\Delta$ and pion degrees of freedom consists of four terms:
\begin{equation}
{\cal L}={\cal L}_{\pi N} + {\cal L}_{\pi\Delta} + {\cal L}_{\pi\Delta N} +
{\cal L}_{\pi\pi}
\end{equation}
The piece of the Lagrangian solely involving pions is given by the standard
expressions of ref. \cite{e} and is not affected by the heavy baryon
transformations to ${\cal O}(\epsilon^2)$.
For the other three components of the Lagrangian we
follow the same philosophy as in section 2. We start from the leading
order relativistic Lagrangians in order to construct the leading order
heavy baryon Lagrangians.  The relativistic $\pi N,\; \pi\Delta$ and
$\pi N\Delta$ Lagrangians
have been given above in Eqs.(\ref{eq:qq},\ref{eq:Lambda}) and
Eq.(\ref{eq:NDelta}).

Given our detailed discussion in section 2, one can now read off the leading
heavy baryon Lagrangians by replacing the relativistic fields with their
corresponding "large" heavy baryon fields
\begin{eqnarray}
          \psi_N (x) & \rightarrow & P_{v}^{+} \; N(x) \; \exp(-i M_{0} \; v
                                   \cdot x) \\
\psi^{i}_{\mu} (x) & \rightarrow & P_{v}^{+} \;
                                   T^{v}_{j} (x) \; \exp(-i M_{0} \; v \cdot x)
                                   \label{eq:Tmu}
\end{eqnarray}
For nucleons, the transition from relativistic $\psi(x)$ to heavy
baryon $N(x)$ was reviewed in section 2.  The prescription given
here to obtain heavy baryon delta degrees of freedom $T^{\mu}(x)$ is
identical to the definition in Eq.(\ref{eq:T}), except that
we now make the choice
\begin{equation}
\exp(-iM_0v\cdot x)
\end{equation}
for the exponential time dependence in order to recover the standard form of
${\cal L}_{\pi N}^{(1)}$ as given in ref. \cite{BKKM92}. We
find then, to lowest order
\begin{eqnarray}
{\cal L}_{\pi N}^{(1)} & = & \bar{N} \left[ i \; v \cdot D \; + \; g_{A}
      \; S \cdot u \right] N \nonumber\\
{\cal L}_{\pi\Delta}^{(1)} & = & - \; \bar{T}^{\mu}_{i} \left[ i \; v\cdot
D^{ij}- \delta^{ij}\Delta_0 + g_{1} \; S \cdot u ^{ij}
     \right] \; g_{\mu\nu} \; T^{\nu}_{j}\nonumber\\
{\cal L}_{\pi N\Delta}^{(1)} & = & g_{\pi N\Delta} \; \left\{ \bar{T}^{\mu}_{i}
                                  \; g_{\mu\alpha} \; w^{\alpha}_{i} \; N \;
                                  + \; \bar{N} \; w^{\alpha\dagger}_{i} \;
                                  g_{\alpha\mu} \; T^{\mu}_{i}
\right\}\label{eq:oo}
\end{eqnarray}
with $\Delta_0=M_{\Delta} - M_{0}$ being the (bare) nucleon-delta mass
splitting.
Aside from this parameter, there are no other changes in the leading order
delta Lagrangian ${\cal L}_{\pi\Delta}^{(1)}$ when compared with
Eq.(\ref{eq:ADelta}) of
section 3, as expected. In contrast to the underlying relativistic lagrangian
Eq.(\ref{eq:NDelta}) the leading order non-relativistic $\pi N\Delta$
Lagrangian
${\cal L}_{\pi N\Delta}^{(1)}$ contains only the on-shell coupling structure
proportional
to $g_{\pi N\Delta}$. This has also been noted by Lucio and Napsuciale
\cite{Napsuciale}.

The non-relativistic constituent quark model in the SU(6) limit suggests
\begin{equation}
g_{\pi N\Delta}^{QM}= g_{A} \frac{6}{5 \sqrt{2}} \approx 1.07 \; .
\end{equation}
However, in contrast to $NN$ and $\Delta\Delta$ transitions
it is known that for the case of the nucleon-delta transition the
quark model estimates are not very reliable \cite{HHM95}. We regard it
therefore as mandatory that all $N\Delta$-transition couplings be fit
from experiment. This said, we have obtained \cite{HHK97c}
a consistent determination\footnote{Note that the leading ``small scale
expansion''
value for $g_{\pi N\Delta}$ given here is different from the one obtained in
the
phenomenological (relativistic) analyses of pion photoproduction in the
$\Delta$(1232)
region ({\em e.g.} \cite{benmerrouche}), which is mainly due to the
non-relativistic
kinematics of the heavy mass method.}
of $g_{\pi N\Delta}$ from a fit to the decay width of the
delta resonance to {\it leading order} in the ``small scale expansion'', which
{\it is} consistent with the SU(6) quark model
\begin{equation}
g_{\pi N\Delta}^{\rm HHK}=1.05 \pm 0.02 \ . \label{eq:gpiNDeltaHHK}
\end{equation}
Details of this determination will be given in a future communication as
we are just concerned with the general structure of the lagrangians in this
paper.

We now move on to consider the ${\cal O}(\epsilon^2)$---$1/M$---corrections
to these leading order lagrangians.

\subsection{1/M Corrected Lagrangians for Nucleon-Delta Interactions}

The leading order relativistic Lagrangians of section 4.2 contain all the
information necessary in order
to construct the first $1/M$ corrections.

At this point it is useful to clarify what is meant by ``$1/M$''
in a system which
contains {\it two} independent mass scales, the (bare) nucleon mass $M_{0}$ and
the (bare) delta mass $M_{\Delta}$. First, we
rewrite all mass dependent expressions into the set $M_{0}$ and $\Delta =
M_{\Delta}-M_{0}$.
In order to establish a systematic chiral power counting for our Lagrangians,
we then take the limit $\Delta << M_{0}$ and expand all quantities as a power
series in $1/(2M_{0})$, with appearances of the ``small'' scale $\Delta$
therefore restricted to numerators\footnote{This is certainly correct for all
terms appearing in our Lagrangians. However, the ``small'' scale $\Delta$ can
enter into the denominator of amplitudes through the spin 3/2 propagator or
spin 3/2 kinematic coefficients. For an example, see our discussion on neutral
pion photoproduction in ref.\cite{HHK96a}}.
This suppression by twice the (bare) mass of the
nucleon is what we mean by the phrase $1/M$. Let us note here, that the choice
$M_{0}$, $\Delta$ as the two independent mass scales is not unique,
but provides us with the most convenient way to match onto the already
existing heavy baryon formalism for pure nucleon degrees of freedom, as for
example laid out in ref.\cite{j}.

For notational simplicity we will again suppress all four-vector and isospin
indices. Aside from the ``large'' spin 3/2
component $T^{\mu}(x)$, we now also have to consider the
other five delta degrees of freedom already discussed in section 3.2.
As before, we denote
these collectively $G^{\mu}(x)$, but now with the mass
parameter being $M_{0}$. Furthermore, using the ``large'' and ``small''
heavy nucleon fields $N,h$ defined in section 2,
we have a complete set of degrees of freedom with which
to rewrite the
relativistic Lagrangians
in term of heavy baryon fields---
\begin{eqnarray}
      {\cal L}_{\pi N} & = & \bar{N} {\bf A}_{N} N + \bar{h} \; {\bf B}_{N} N +
                      \bar{N} \; \gamma_{0} {\bf B}_{N}^{\dagger} \gamma_{0} \;
                      h - \bar{h} \; {\bf C}_{N} \; h  \nonumber\\
  {\cal L}_{\pi\Delta} & = & \bar{T} {\bf A}_{\Delta} T + \bar{G} \; {\bf
B}_{\Delta}
                      T + \bar{T} \; \gamma_{0} {\bf B}_{\Delta}^{\dagger}
                      \gamma_{0} \; G - \bar{G} \; {\bf C}_{\Delta} \; G
                    \nonumber\\
{\cal L}_{\pi\Delta N} & = & \bar{T} {\bf A}_{\Delta N} N + \bar{G} \;
                      {\bf B}_{\Delta N} N + \bar{T} \; \gamma_{0}
                      {\bf D}_{N\Delta}^{\dagger} \gamma_{0} \; h + \bar{G} \;
                      \gamma_{0} {\bf C}_{N\Delta}^{\dagger} \gamma_{0} \; h
                      \nonumber \\
                & +  & \bar{N} \; \gamma_{0} {\bf A}_{\Delta N}^{\dagger}
                      \gamma_{0} \; T + \bar{N} \; \gamma_{0}
                      {\bf B}_{\Delta N}^{\dagger} \gamma_{0} \; G + \bar{h} \;
                      {\bf D}_{N\Delta} T + \bar{h} \; {\bf C}_{N\Delta}
                      \; G \label{eq:pp}
\end{eqnarray}
If one is only interested in the ${\cal O}(\epsilon^2)$ $1/M$ corrections,
several
useful simplifications occur in these forms:
\begin{itemize}

\item [i)] all matrices ${\bf A}_{X}$, ${\bf B}_{X}$, ${\bf C}_{X}$,
${\bf D}_{X}$
are only needed to ${\cal O}(\epsilon)$. The structures depending on
${\bf A}_{X}^{(1)}$ matrices are then given by the ${\cal O}(\epsilon)$
Lagrangians of Eq.(\ref{eq:oo}).

\item [ii)] the matrix ${\bf D}_{N\Delta}^{(1)}$ is found to be identically
zero.

\item [iii)] all terms in Eq.(\ref{eq:pp}) that couple $G^{\mu}(x)$
and $h(x)$ degrees of freedom can be discarded in this section as they only
start contributing at ${\cal O}(\epsilon^{4})$.

\end{itemize}
In order to obtain the Lagrangians at ${\cal O}(\epsilon^{2})$, one can
therefore
simply use the
variable shift of ref. \cite{BKKM92} for the ``small" nucleon component
\begin{equation}
h \rightarrow h' + {\bf C}_{N}^{-1} \; {\bf B}_{N}^{(1)} \; N
\end{equation}
and then integrate out the $h'(x)$  degrees of freedom as in section
2.
Similarly, we introduce an analogous shift for the $G^{\mu}(x)$ fields
\begin{equation}
G \rightarrow G' + {\bf C}_{\Delta}^{-1} \; {\bf B}_{\Delta}^{(1)} \;
T + {\bf C}_{\Delta}^{-1} \; {\bf B}_{\Delta N}^{(1)} \; N,
\label{eq:Gprime3}
\end{equation}
in order to write our Lagrangians exclusively in terms of the ``large"
fields $T^{\mu}(x)$ and $N(x)$. Note that the variable change in
Eq.(\ref{eq:Gprime3}) is more complex than the corresponding one in
section 3
due to the addition of spin 1/2 fields.  Finally,
integrating out the $G'(x)$ degrees of freedom\footnote{Any possible effects by
the
associated determinant can again only start showing up at ${\cal
O}(\epsilon^3)$
and are therefore not considered here.}
leaves us with the following formal expressions
for the ${\cal O}(\epsilon^2)$ $1/M$ Lagrangians:
\begin{eqnarray}
 {\cal L}_{\pi N}^{(2)} & = & \bar{N} \left[ \gamma_{0}
                                {\bf B}_{N}^{{(1)} \dagger} \gamma_{0}
                                {\bf C}_{N}^{-1} {\bf B}_{N}^{(1)} +
                                \gamma_{0} {\bf B}_{\Delta N}^{{(1)}
                                \dagger} \gamma_{0} {\bf C}_{\Delta}^{-1}
                                {\bf B}_{\Delta N}^{(1)} \right] N
                             \nonumber\\
 {\cal L}_{\pi\Delta}^{(2)} & = & \bar{T} \gamma_{0} {\bf B}_{\Delta}^{{(1)}
                                \dagger} \gamma_{0} {\bf C}_{\Delta}^{-1}
                                {\bf B}_{\Delta}^{(1)} T
                                \nonumber\\
 {\cal L}_{\pi\Delta N}^{(2)} & = & \bar{T} \gamma_{0} {\bf B}_{\Delta}^{{(1)}
                                \dagger} \gamma_{0} {\bf C}_{\Delta}^{-1}
                                {\bf B}_{\Delta N}^{(1)} N +
                                \bar{N} \gamma_{0}
                                {\bf B}_{\Delta N}^{{(1)} \dagger}
                                \gamma_{0} {\bf C}_{\Delta}^{-1}
                                {\bf B}_{\Delta}^{(1)} T
\end{eqnarray}
We begin our discussion with the ${\cal O}(\epsilon^2)$ $1/M$ Lagrangian for
nucleons
${\cal L}^{(2)}_{\pi N}$. Aside from the well known piece $\gamma_{0}
{\bf B}_{N}^{{(1)} \dagger} \gamma_{0} {\bf C}_{N}^{-1}
{\bf B}_{N}^{(1)}$ of Eq.(\ref{eq:1mnn}) in section 2.2, we encounter the
new contributions
$\gamma_{0} {\bf B}_{\Delta N}^{{(1)} \dagger} \gamma_{0}
{\bf C}_{\Delta}^{-1} {\bf B}_{\Delta N}^{(1)}$ which are proportional
to $g_{\pi N\Delta}^{\hspace{1mm} 2}$. As one can very clearly see from the
matrix structure, these terms
arise from intermediate $G^{\mu} (x)$ states that were
integrated out, resulting in new $1/M$ suppressed two-pion vertices in the
nucleon Lagrangian. Starting from the relativistic nucleon-delta transition
Lagrangian ${\cal L}_{\pi N\Delta}$
and translating into the heavy baryon formalism
as outlined in section 3 for the case of the pure spin 3/2 Lagrangian, we find
the explicit representation for the matrix
${\bf B}_{\Delta N}^{(1)}$
\begin{eqnarray}
& & \nonumber \\
{\bf B}_{\Delta N, \mu}^{{(1)}i} &=& \left(
                          \begin{array}{c}
                          0 \\ \\
                          g_{\pi N\Delta} \; (1 + 3 z_0) \; P_{+} \;
                          _{(11)}P^{1/2}_{\mu\nu} \; P_{+} \; w^{i\nu}
                          \\ \\
                          g_{\pi N\Delta} \; \sqrt{3} \; z_0 \; P_{-} \;
                          _{(21)}P^{1/2}_{\mu\nu} \; P_{+} \; w^{i\nu}
                          \\ \\
                          g_{\pi N\Delta} \; \sqrt{3} \; z_0 \; P_{-} \;
                          _{(12)}P^{1/2}_{\mu\nu} \; P_{+} \; w^{i\nu}
                          \\ \\
                          g_{\pi N\Delta} \; (1 + z_0) \; P_{+} \;
                          _{(22)}P^{1/2}_{\mu\nu} \; P_{+} \; w^{i\nu}
                          \\
                          \end{array}
                          \right) \label{eq:B1DeltaN} \\
& & \nonumber
\end{eqnarray}

Furthermore, we need ${\bf C}_{\Delta}^{-1}$ at ${\cal O}(1/M$) which is
given in Appendix C. After some algebra one then arrives at
\begin{eqnarray}
                    &   & \nonumber \\
{\cal L}_{\pi N}^{(2)} & = & \bar{N} \; \gamma_{0} {\bf B}_{N}^{{(1)} \dagger}
                          \gamma_{0} \; {\bf C}_{N}^{-1} \;
                          {\bf B}_{N}^{(1)} \; N - \frac{1}{2 M_{0}} \; \bar{N}
\left[
                          g_{\pi\pi NN}^{(2a)} \; w^{i\dagger} \cdot S \;
                          \xi^{ij}_{3/2} \;S \cdot w^{j} \right. \nonumber \\
                    &   & \nonumber \\
                    &   & \left. +
                          g_{\pi\pi NN}^{(2b)} \; w^{i\dagger} \cdot v \;
                          \xi^{ij}_{3/2} \; v \cdot w^{j} \;
                          \right] N  \nonumber \\
                    &   & \label{eq:piN1/M}
\end{eqnarray}
One realizes that the two new structures start out as $\pi\pi NN$ coupling
terms,
as expected from the leading order $\pi N\Delta$ lagrangian
Eq.(\ref{eq:NDelta}).
The new two-pion coupling constants $g_{\pi\pi NN}^{(2a)}$,
$g_{\pi\pi NN}^{(2b)}$ are given in terms of the ${\cal O}(\epsilon)$
nucleon-delta
coupling constants $g_{\pi N\Delta}, z_0$ and are likely
to play a significant role in low-energy pion-nucleon scattering and
near-threshold two-pion production amplitudes. We note that
these two coupling constants are related to the ${\cal O}(\epsilon^{2})$
counterterms
$c_{2}, c_{3}, c_{4}$ of ref.\cite{j}. A new analysis of the anatomy of
these counterterms is called for. Here we only define the general structures
\begin{eqnarray}
g_{\pi\pi NN}^{(2a)} & = & g_{\pi N\Delta}^{\hspace{1mm} 2} \; \frac{4}{3} \;
                             ( 1 + 8 z_0 + 12 z^{2}_0 )
                             \nonumber \\
g_{\pi\pi NN}^{(2b)} & = & g_{\pi N\Delta}^{\hspace{1mm} 2} \; \frac{1}{3} \;
                             ( 5 - 8 z_0 - 4 z^{2}_0 )
                             \label{eq:g2b}
\end{eqnarray}
We now move on to the ${\cal O}(\epsilon^2)$ $1/M$ Lagrangian for deltas
${\cal L}^{(2)}_{\pi\Delta}$. Noting that ${\bf B}_{\Delta}^{(1)} =
{\cal B}_{\Delta}^{(1)}$ and employing
${\bf C}_{\Delta}^{-1}$ from Appendix C, we determine
\begin{eqnarray}
{\cal L}_{\pi \Delta}^{(2)} = \frac{1}{2 M_{0}} \; \bar{T}^{\mu}_{i} (x) \;
& \mbox{\bf \Large\{ } & \left[ \; D_{\alpha}^{ik} D_{\beta}^{kj}
                         \; g^{\alpha\beta} - v \cdot D^{ik} v
                         \cdot D^{kj} \right] \; g_{\mu\nu}  \nonumber \\
                   & + & g_{1} \; i \left( S \cdot D^{ik} v
                         \cdot u^{kj} + v \cdot u^{ik}S \cdot D ^{kj}
                         \right) \; g_{\mu\nu} \nonumber \\
                   & - & \left[ S^{\alpha} , S^{\beta} \right] \left(
                         D_{\alpha}^{ik} D_{\beta}^{kj} - D_{\beta}^{ik}
                         D_{\alpha}^{kj} \right) \; g_{\mu\nu}
                         \nonumber \\
                   & + & \frac{g_{1}^{2}}{4} v \cdot u^{ik}
                         v \cdot u^{kj} \; g_{\mu\nu} \nonumber \\
                   & - & g_{\pi\pi\Delta\Delta} \; u_{\mu}^{ik}
                         u_{\nu}^{kj} \; \mbox{\bf \Large\} } T^{\nu}_{j} (x)
                         \label{eq:piDelta1/M}
\end{eqnarray}
with
\begin{equation}
g_{\pi\pi\Delta\Delta} =
\frac{1}{3} \; \left( \; g_{1}^{2} + 4 \; g_{1} \; g_{2} + 3
                   \; g_{2}^{2} \; \right) \label{eq:gpipiDelta}
\end{equation}
This form is identical to that of Eq.(\ref{eq:ss}) except for the change
$1/2M_\Delta\rightarrow 1/2M_0$ and the associated physics has
already been discussed in section
3.4.  The $1/M$
suppressed two-pion delta-delta coupling $g_{\pi\pi\Delta\Delta}$ is
in principle
determined in terms of the ${\cal O}(\epsilon)$
coupling constants $g_{1}$, $g_{2}$.
At present, however, we have no information about the
off-shell coupling constant $g_{2}$ and therefore treat $g_{\pi\pi\Delta}$
as a low energy constant of ${\cal O}(\epsilon^{2})$ to be fitted from future
experiment.  Theoretically speaking, it nevertheless has a
status different from the ${\cal O}(\epsilon^{2})$ counterterms of section 5.

Finally, we turn to the ${\cal O}(\epsilon^{2})$ $1/M$ nucleon-delta transition
Lagrangian ${\cal L}_{\pi N\Delta}$. With ${\bf B}_{\Delta}^{(1)}$
and ${\bf B}_{\Delta N}^{(1)}$ from above and
$({\bf C}_{\Delta}^{-1})^{(0)}$ from Appendix C we find
\begin{eqnarray}
{\cal L}_{\pi\Delta N}^{(2)} &=& \frac{-1}{2M_{0}} \; \bar{T}^{\mu}_{i} (x)
\left[ h_{\pi\pi N\Delta} \; u_{\mu}^{ij}\xi^{jk}_{3/2}\;S \cdot
                       w^{k}
                  + 2 \; g_{\pi N\Delta} \; i D_{\mu}^{ij}\xi^{jk}_{3/2}\;
                       v \cdot w^{k} \right]
                       N(x)+h.c. \nonumber \\
& & \label{eq:piDeltaN1/M}
\end{eqnarray}
with
\begin{equation}
h_{\pi\pi N\Delta} = \frac{2}{3} \; g_{\pi N\Delta}
                     \left( \; g_{1} + 2 \; g_{2} + 4 \; g_{1} \; z_0
                     + 6 \; g_{2} \; z_0 \; \right) \label{eq:hpiNDelta}
\end{equation}
The first term in this ${\cal O}(\epsilon^{2})$ nucleon-delta transition
Lagrangian
starts out as a two-pion vertex. In fact, this is the lowest order vertex of
this kind, as the corresponding ${\cal O}(\epsilon)$ Lagrangian ${\cal
L}^{(1)}_{\pi \Delta N}$ Eq.(\ref{eq:NDelta})
contains only vertices involving odd number of pions.  The new coupling
constant
$h_{\pi\pi N\Delta}$ is of the same type as $g_{\pi\pi\Delta\Delta}$.
In principle it is fixed in terms of the
${\cal O}(\epsilon)$
coupling constants $g_{\pi N\Delta}, z_0, g_{1}, g_{2}$, but again in practice
we are
ignorant about the size of $g_{2}$ and therefore must treat it as a free
parameter of ${\cal O}(\epsilon^{2})$, at least for now. The second term in
Eq.(\ref{eq:piDeltaN1/M}) is completely determined by the ${\cal O}(\epsilon)$
coupling
constant $g_{\pi N\Delta}$ and starts out as a $1/M$ suppressed one-pion
vertex, which is non-vanishing even at threshold. As we have shown in
ref.\cite{HHK96a}, this structure contributes to neutral pion
photoproduction at order $\epsilon^3$.

This concludes our discussion on the $1/M$ corrections at ${\cal
O}(\epsilon^{2})$.
We next introduce a general formalism which allows construction of
terms to arbitrary order in the chiral expansion.

\subsection{The Small Scale Expansion to All Orders}

We begin our all orders discussion from the set of nucleon
and delta Lagrangians that have
already been translated into the heavy baryon formalism, using the
notation introduced above to illustrate our method.
As has been done in section 3, we will use path integral methods to
integrate out ``small'' components of the Dirac wavefunctions.
The first step is to decouple the $h(x)$ and $G^{\mu}(x)$ degrees of freedom
via the change of variables
\begin{equation}
G \rightarrow G' + {\bf C}_{\Delta}^{-1} \; \gamma_0
{\bf C}_{N\Delta}^{\dagger}\gamma_0 \; h,
\end{equation}
resulting in the new set of Lagrangians
\begin{eqnarray}
 {\cal L}_{\pi N}' & = & \bar{N} \; {\bf A}_{N} \; N + \bar{h} \;
                       {\bf \tilde{B}}_{N} \; N + \bar{N} \; \gamma_{0}
                       {\bf \tilde{B}}_{N}^{\dagger} \gamma_{0} \; h - \bar{h}
                       \; {\bf \tilde{C}}_{N} \; h \nonumber\\
  {\cal L}_{\pi\Delta}' & = & \bar{T} \; {\bf A}_{\Delta} \; T + \bar{G}' \;
                       {\bf B}_{\Delta} \; T + \bar{T} \; \gamma_{0}
                       {\bf B}_{\Delta}^{\dagger} \gamma_{0} \; G' - \bar{G}'
                       \; {\bf C}_{\Delta} \; G' \nonumber\\
  {\cal L}_{\pi\Delta N}' & = & \bar{T} \; {\bf A}_{\Delta N} \; N + \bar{G}'
\;
                       {\bf B}_{\Delta N} \; N + \bar{T} \; \gamma_{0}
                       {\bf \tilde{D}}_{N\Delta}^{\dagger} \gamma_{0}
                       \; h \nonumber\\
                       &+&\bar{h}\tilde{\bf{D}}_{N\Delta}T+\bar{N}\gamma_0
                       \bf{B}^\dagger_{\Delta N}\gamma_0G'
                       +\bar{N}\gamma_0{\bf A}_{\Delta N}^\dagger\gamma_0T
\end{eqnarray}
with
\begin{eqnarray}
{\bf \tilde{B}}_{N}       & = & {\bf B}_{N} + {\bf C}_{N\Delta} \;
                                C_{\Delta}^{-1} \; {\bf B}_{\Delta N}
\nonumber\\
{\bf \tilde{C}}_{N}       & = & {\bf C}_{N} - {\bf C}_{N\Delta} \;
                                C_{\Delta}^{-1} \;
                                {\bf C}_{N\Delta}^{\dagger} \nonumber\\
{\bf \tilde{D}}_{N\Delta} & = & {\bf D}_{N\Delta} + {\bf C}_{N\Delta} \;
                                C_{\Delta}^{-1} \; {\bf B}_{\Delta}
\end{eqnarray}
The next step involves decoupling the ``small" fields $h(x),
{G^{\mu}}'(x)$ from
the ``large" fields $N(x), T^{\mu} (x)$---we shift the fields
according to
\begin{eqnarray}
G' & \rightarrow & G'' + {\bf C}_{\Delta}^{-1} \; {\bf B}_{\Delta} \; T +
                   {\bf C}_{\Delta}^{-1} \; {\bf B}_{\Delta N} \; N \nonumber\\
h  & \rightarrow & h' + {\bf \tilde{C}}_{N}^{-1} \; {\bf \tilde{B}}_{N} \;
                   N + {\bf \tilde{C}}_{N}^{-1} \;
                   {\bf \tilde{D}}_{N\Delta} \; T
\end{eqnarray}
and then integrate out the quantities $h'(x), {G^{\mu}}'' (x)$.
The resulting Lagrangians, which no longer involve couplings to
$h',{G^{\mu}}''$ read
\begin{eqnarray}
{\bf \tilde{L}}_{\pi N} & = & \bar{N} {\bf A}_{N} N + \bar{N} \left[ \;
                                    \gamma_{0} {\bf \tilde{B}}_{N}^{\dagger}
                                    \gamma_{0} \; {\bf \tilde{C}}_{N}^{-1}
                                    \; {\bf \tilde{B}}_{N} + \gamma_{0}
                                    {\bf B}_{\Delta N}^{\dagger} \gamma_{0} \;
                                    {\bf C}_{\Delta}^{-1} \;
                                    {\bf B}_{\Delta N} \right] N
                                     \nonumber\\
{\bf \tilde{L}}_{\pi\Delta} & = & \bar{T} {\bf A}_{\Delta} T + \bar{T} \left[
                                    \; \gamma_{0} {\bf B}_{\Delta}^{\dagger}
                                    \gamma_{0} \; {\bf C}_{\Delta}^{-1} \;
                                    {\bf B}_{\Delta} + \gamma_{0}
                                    {\bf \tilde{D}}_{N\Delta}^{\dagger}
                                    \gamma_{0} \; {\bf \tilde{C}}_{N}^{-1}
                                    \; {\bf \tilde{D}}_{N\Delta} \; \right] T
                                    \label{eq:allDelta} \nonumber\\
{\bf \tilde{L}}_{\pi\Delta N} & = & \bar{T} {\bf A}_{\Delta N} N + \bar{T}
                                    \left[ \; \gamma_{0}
                                    {\bf \tilde{D}}_{N\Delta}^{\dagger}
                                    \gamma_{0} \; {\bf \tilde{C}}_{N}^{-1}
                                    \; {\bf \tilde{B}}_{N} + \gamma_{0}
                                    {\bf B}_{\Delta}^{\dagger} \gamma_{0} \;
                                    {\bf C}_{\Delta}^{-1} \;
                                    {\bf B}_{\Delta N} \right] N\nonumber\\
           & +  & \bar{N} \; \gamma_{0}
                                {\bf A}_{\Delta N}^{\dagger} \gamma_{0} \; T
                                    + \bar{N} \left[ \; \gamma_{0}
                                {\bf \tilde{B}}_{N}^{\dagger} \gamma_{0} \;
                                    {\bf \tilde{C}}_{N}^{-1} \;
                                    {\bf \tilde{D}}_{N\Delta} + \gamma_{0}
                                    {\bf B}_{\Delta N}^{\dagger} \gamma_{0} \;
                                    {\bf C}_{\Delta}^{-1} \;
                                    {\bf B}_{\Delta} \right] T
\nonumber\\
\quad
\end{eqnarray}
One can then read off the desired $1/M$ corrections to arbitrary
order from the matrix
expressions in the square brackets.  In
particular, one can verify that the ${\cal O}(\epsilon^{2})$ $1/M$
Lagrangians of section 4.3 are correct and complete.

Finally, we would like to stress that matrices ${\bf A}_{X}$, ${\bf
B}_{X}$, and ${\bf C}_{X}$ are derived from the corresponding
relativistic lagrangians. The corresponding expressions have been
given in previous sections mostly only to leading order. However, when
going to higher orders in the expansion,
it is necessary to augment such terms by counterterm Lagrangians.

In section 5 we construct the counterterm Lagrangians of
${\cal O}(\epsilon^{2})$ in the small scale expansion.

\section{Counterterms}

Before undertaking applications of the formalism developed above there
remains one important step. Working in an effective field theory
framework, it is mandatory to include all possible local terms allowed
by the symmetry requirement. These so called counterterms appear at
each order $\epsilon^n$ of the low energy expansion, starting at order
$\epsilon^2$.
Also, when loop contributions are taken into
account, various divergences will arise which must be absorbed into
the counterterm component of the effective action.
For example, in
HBChPT the most general structures have been given up to order
$p^3$ in \cite{EM96} for the case of SU(2).
Here we explicitly construct the  next-to-leading order
heavy baryon $NN$-, $N\Delta$- and $\Delta\Delta$-counterterm lagrangians,
starting from the corresponding relativistic chiral baryon lagrangians.
The general methods described here can be generalized for the
construction of arbitrary higher order counterterm lagrangians
in the three sectors of the theory.  However, we shall quote the
results explicitly only to ${\cal O}(\epsilon^2)$.  Also, because of
space limitations we shall be content merely to sketch the derivation
of these results.  A more complete discussion is available in
ref.\cite{HThes97}.

The basic procedure which we utilize is to construct the most general
form of the relativistic Lagrangian in the sector being considered.
We refer to \cite{GSS88,Krause} for examples from the spin 1/2 sector.
In performing this task, it should be kept in mind that
the only information we are using about the
strong interactions at low energies are the symmetries which the
effective lagrangian and the meson/baryon fields have to obey. In
particular, these are symmetries under parity transformations,
charge conjugation, hermitean conjugation and overall Lorentz-invariance,
as well as invariance
under chiral vector and chiral axial-vector transformations. Violation of
one of these symmetries is the only justification to omit a possible
structure in the counterterm lagrangians to the order we are
working. Our approach is
to implement these symmetries on the level of the relativistic lagrangians
and then to perform the (non-relativistic) 1/M expansion.

\subsection{Building Blocks and Chiral Counting Rules}

One begins the process by itemizing the various building blocks from
which to form such a Lagrangian. In addition to the already defined
structures we employ:
\begin{eqnarray}
\chi_\pm       &=& 2 B \left[ u^\dagger
                   \left( {\bf s} + i {\bf \tilde{p}} \right)
                   u^\dagger \pm u
                   \left( {\bf s} + i {\bf \tilde{p}} \right)^\dagger
                   u \right] \equiv \tau^i \; \chi_{\pm}^i \; ,
\nonumber \\
\chi^{(s)}_\pm &=& \frac{1}{2} \; Tr \left( \chi_\pm \right) \; ,
\nonumber \\
f_{\mu\nu}^\pm &=& u^\dagger F_{\mu\nu}^R u \pm u F_{\mu\nu}^L
                   u^\dagger \equiv \tau^i \; f_{\pm\mu\nu}^i\; ,
\nonumber \\
F_{\mu\nu}^X   &=& \partial_\mu F_{\nu}^X - \partial_\nu F_{\mu}^X - i
                   \left[ F_{\mu}^X , F_{\nu}^X \right] ; \;\; X = L
                   , R \; , \nonumber \\
F_{\mu}^R      &=& {\bf v_\mu} + {\bf a_\mu} \;, \hspace{5mm} F_{\mu}^L \; =
                   \; {\bf v_\mu} - {\bf a_\mu} \; , \nonumber \\
v_{\mu\nu}^{(s)}&=& \partial_\mu v_{\nu}^{(s)} - \partial_\nu
                    v_{\mu}^{(s)} \; , \nonumber \\
w_{\mu\nu}^i    &=& \frac{1}{2} \; Tr \left( \tau^i \left[ D_\mu , u_\nu
                          \right] \right) \label{eq:wmni}\; .
\end{eqnarray}
Here ${\bf s} \; ( {\bf \tilde{p}} )$ denote an external scalar (pseudoscalar)
field,
whereas ${\bf v_\mu} \; ( {\bf a_\mu} )$ correspond to an external isovector
vector (axial-vector) field. Following ref.\cite{EM96} we have also defined
$v_{\mu}^{(s)}$ denoting an external isoscalar vector field, which we need
for processes involving external photons.

Having defined the building blocks of our lagrangians, we note that
each has certain properties under chiral power counting.  Thus, for
example, forms such as
\begin{eqnarray}
M_0, \; M_\Delta, \;
\psi_N , \; \psi_{\mu}^j, \; D_\mu \psi_N, \; D_{\mu}^{ij} \psi_{\nu}^j, \; U
\end{eqnarray}
count as ${\cal O}(\epsilon^0)$, while others, {\it e.g.}
\begin{eqnarray}
\nabla_\mu U, \; u_\mu, \; w_{\mu}^i, \; \Delta, \;
\left( i \not{\!\!D} - M_0 \right) \psi_N, \;
\left( i \not{\!\!D}^{ij} - M_\Delta \right) \psi_{\mu}^{j}
\end{eqnarray}
contribute at ${\cal O}(\epsilon^1)$, while still others
\begin{eqnarray}
m_q , \; \chi_\pm, \; \chi^{(s)}_\pm, \; \chi_{\pm}^i , \;
f_{\mu\nu}^\pm, \; v_{\mu\nu}^{(s)}, \; f_{\pm\mu\nu}^i, \;
\left[ D_\mu , u_\nu \right], \; w_{\mu\nu}^i
\end{eqnarray}
count as ${\cal O}(\epsilon^2)$.
\footnote{We work in the framework of standard ChPT and not the
Generalized ChPT of J.~Stern et al. \cite{SSF93}.}
Using these forms then one can easily construct Lorentz-invariant
relativistic Lagrangians which begin at any particular order of chiral
counting.

\subsection{Transformation Rules}

The ways by which to enforce parity, charge conjugation
invariance and hermiticity in relativistic chiral lagrangians for baryons
are well-known and do not need repeating
here. For details we refer to \cite{Krause,HThes97}. However, less
familiar perhaps are the strictures associated
with invariance under chiral rotations under which the structures
defined above transform according to \cite{e,GSS88}
\begin{eqnarray}
U=u^2         &\rightarrow& V_R \; U \; V_{L}^\dagger \nonumber \\
u             &\rightarrow& V_R \; u \; h^\dagger = h \; u \; V_{L}^\dagger
                            \label{eq:compensator} \nonumber \\
\nabla_\mu U  &\rightarrow& V_R \; \nabla_\mu U \; V_{L}^\dagger
\nonumber \\
\psi_N        &\rightarrow& h \; \psi_N \nonumber \\
X             &\rightarrow& h \; X \; h^\dagger \;\;\; \mbox{with} \quad
 X= D_\mu, \; u_\mu, \; \chi_\pm, \; \chi_{\pm}^{(s)}, \; v_{\mu\nu}^{(s)},
\; f_{\mu\nu}^\pm \; .
\end{eqnarray}
Here the operators $V_L \; (V_R)$ denote  chiral rotations among the
left- (right-) handed quarks of the underlying QCD lagrangian, whereas
$h=h(V_R, V_L, \pi)$ corresponds to the ``compensator field''
\cite{CCWZ69,Georgi} of the nonlinear
chiral representation, defined in Eq.(\ref{eq:compensator}).

Defining chiral vector $V_V$ and chiral axial $V_A$ transformations via
\begin{eqnarray}
V_V(\beta^i) = V_R + V_L \; , \quad
V_A(\alpha^i) = V_R - V_L \; ,
\end{eqnarray}
we can give a representation of the compensator field $h$ for infinitesimal
chiral SU(2) transformations:
\begin{equation}
h(\alpha^i,\beta^j,\pi^k) = 1 + i \frac{\tau \cdot \beta}{2} - \frac{i}{2F_0}
                            \; \alpha^i \pi^k \epsilon^{ikl} \; \frac{\tau^l}
                            {2} + {\cal O}({\bf \alpha^2},{\bf \beta^2},{\bf
                            \pi^2}) \; . \label{eq:infcomp}
\end{equation}
Here the three (infinitesimal) rotation angles $\alpha^i$ correspond to chiral
axial
rotations. This symmetry sector of the hadronic theory is spontaneously broken
at low energies, resulting in the emission and
absorption of three associated Goldstone bosons $\pi^k$ (pions) with
associated decay constant $F_0$, as is
evident from Eq.(\ref{eq:infcomp}).

For the chiral objects with explicit isospin index we use the transformation
law
\begin{eqnarray}
Y^i&\rightarrow& h^{ij} \; Y^j \; h^\dagger \; , \quad
 Y^i= w_{\mu}^i, \;
                 \chi_{\pm}^i, \;
                  f_{\pm\mu\nu}^i, \; w_{\mu\nu}^i  \; ,
\end{eqnarray}
with the chiral response matrix\footnote{Ellis and Tang have given an
expression
for the spin 3/2 isospin 3/2 compensator field $h^{ij}$ without going back to
an
infinitesimal representation. \cite{ET96}}
\begin{eqnarray}
h^{ij} =\left[ \delta^{ij} + \left( \frac{i}{2} \delta^{ij}
                           \tau^k - \epsilon^{ijk} \right) \beta^k + \frac{1}{
                           2 F_0} \left( i \delta^{ij} \epsilon^{abk} \frac{
                           \tau^a}{2} \pi^b + \pi^i \delta^{jk} - \delta^{ik}
                           \pi^j \right) \alpha^k \right] \; . \nonumber
\end{eqnarray}
Finally, the covariant derivative acting on the I=3/2 field and the spin
3/2 isospin 3/2 field itself transform as
\begin{eqnarray}
\psi_{\mu}^i &\rightarrow& h^{ij} \; \psi_{\mu}^j \nonumber \\
D_{\nu}^{ij} \psi_{\mu}^j &\rightarrow& h^{ia} D_{\nu}^{ab} \psi_{\mu}^b
\; .
\end{eqnarray}

\subsection{${\cal O}(\epsilon^2)$ Counterterm Lagrangians}

Having established the behavior of building blocks under chiral
rotations, we now construct the appropriate relativistic lagrangians.
We start from a set of relativistic interaction structures in each sector
($NN$, $\Delta N$, $\Delta\Delta$) which conserve parity and are
Lorentz invariant. In the
second step we match the free isospin-indices in all possible
combinations.
The final step consists of ensuring hermiticity and C-invariance.
We note the following simplifications of possible structures:

a) All O($\epsilon^2$) {\em relativistic} lagrangians
can be written free of the
equations of motions (EOM) via field redefinitions. EOM terms can only
start appearing in O($\epsilon^3$) lagrangians.

b) All Rarita-Schwinger fields are accompanied by a ``Theta-tensor''
Eq. (\ref{eq:Theta}) which contains a free parameter governing the
coupling to spin 3/2 off-shell degrees of freedom. Throughout this work we
assume that {\em all} off-shell dependent structures involving relativistic
spin 3/2 fields can be rewritten into the ``Theta-tensor'' form, using
Dirac identities and field redefinitions.
All structures that involve the dot-product of a Dirac matrix and a
Rarita-Schwinger spinor ($\gamma_\mu \psi^\mu$) are therefore accounted
for solely by ``Theta-tensors''.

c) One can always move a (covariant) derivative onto the remaining
fields in the relativistic structure via integration by
parts, as the whole lagrangian is only unique up to a total
derivative.

d) As a consequence of a) and c) there are no
structures $\not\!\!{D}, \; \sigma_{\mu\nu}D^\nu$ acting on a baryon field
in the O($\epsilon^2$) lagrangians. We have also assumed that the
O($\epsilon$) $N\Delta$ lagrangian Eq.(\ref{eq:NDelta})
can be written free of $\not\!\!{D}$
structures (EOM terms). Therefore the relativistic O($\epsilon^2$) $N\Delta$
lagrangians also do not contain the structure
$\bar{\psi}_{i}^\mu ( {\not\!\!{D}} w_{\mu}^i ) \psi_N + h.c.$.

e) There are no structures of the type $D_\mu
\psi^\mu$ at O($\epsilon^2$), as this relation is a consequence of the
O($\epsilon$)
EOM
for the relativistic spin 3/2 field in combination with
assumption b).

f) For the same reasoning as in e) we have
also omitted the structure $\bar{\psi}_{i}^\mu {\not\!\!{w}^i} D_\mu
\psi_N + h.c.$ in the O($\epsilon$) $N\Delta$ lagrangian Eq.(\ref{eq:NDelta}).
The O($\epsilon^2$)
contribution contained in this term has been accounted for
after an integration by parts.

The structures proportional to ${\cal A}_{N}^{(2)}$ which are consistent with
invariance under charge- and hermitean
conjugation can then be written in the form \cite{BKKM92}
\begin{eqnarray}
{\cal L}_{N}^{(2)} &=& \bar{N}\mbox{\LARGE \{} c_1 2 \chi^{(s)}_+
+ c_2 \left( v \cdot u \right)^2 + c_3 u \cdot u + c_4 \left[ S_\mu , S_\nu
\right] u^\mu u^\nu \label{eq:ctN2} \label{eq:ctNN} \\
& & \nonumber \\
& & \hspace{.5cm} + c_5 \left[ \chi_+ - \chi^{(s)}_+
\right] - \frac{i}{4 M_0} \left[ S^\mu , S^\nu \right]_- \left[ c_6
f_{\mu\nu}^+
+ 2 c_7 v_{\mu\nu}^{(s)} \right] \mbox{\LARGE \}}N \; . \nonumber
\end{eqnarray}
We note explicitly that Eq.(\ref{eq:ctNN}) contains only the counterterms
of the O($\epsilon^2$) nucleon-nucleon lagrangian! In order to obtain the
complete
O($\epsilon^2$) spin 1/2 lagrangian one still has to add the leading 1/M
$NN$-structures which have been calculated in chapter 4.3. We now give a
brief discussion of some of the physics contained in the counterterm
lagrangian:

The first structure in Eq.(\ref{eq:ctNN}) contains the isoscalar
component of the quark-mass
contribution to the mass of the nucleon. The terms proportional to $c_2$,
$c_3$, $c_4$ start out as $NN\pi\pi$ vertices, which are new structures
beyond the $NN\pi\pi$-vertex (Weinberg term) incorporated in the chiral
connection $D_\mu$ of the leading order $NN$-lagrangian. The fifth term
represents the isovector component (i.e. $\sim \; m_u - m_d$) of the
quark-mass contribution to the mass of the nucleon and therefore vanishes in
the SU(2) isospin symmetry limit. Finally, the structure $c_6$ ($c_7$) can be
related to the isovector (isoscalar) anomalous magnetic moment of the nucleon.

In the same way one can treat the $\Delta N$ and $\Delta\Delta$
sectors of the theory. Once more we only list the
results---details can be found in ref.\cite{HThes97}.
\begin{eqnarray}
{\cal L}_{\Delta N}^{(2)} &=& \bar{T}^{\mu}_i \; \frac{1}{2M_0}
\left[ b_1 \;
i f_{+\mu\nu}^{\; i} S^\nu + b_2 \; i f_{-\mu\nu}^{\; i} v^\nu + b_3 \;
i w_{\mu\nu}^i v^\nu \right. \nonumber \\
&& \nonumber \\
&& \phantom{\bar{T}^{\mu}_i \;\frac{1}{2M_0}  } \left.
   + b_4 \; w_{\mu}^i S \cdot u  + b_5 \; u_\mu S \cdot w^i
   \right] N + h.c. \label{eq:DN2}
\end{eqnarray}
\newline
The physics incorporated in these five resulting structures can very easily
be interpreted. The first term in the lagrangian provides the M1
isovector\footnote{As expected, there is no structure in Eq.(\ref{eq:DN2})
depending on an external isoscalar vector field, as it cannot contribute to
the $\Delta I = 1$ transition of the $N\Delta$ system.} $\gamma N\Delta$
transition
moment, whereas the second term can be determined from one of the axial-vector
nucleon-delta transition form factors at zero four-momentum
transfer. The structure
proportional $b_3$ represents a new $\pi N\Delta$ coupling, that is independent
of the leading coupling $g_{\pi N\Delta}$. The remaining two terms start out
as $N\Delta\pi\pi$ couplings, which are forbidden in the leading order
lagrangian. Again, in order to obtain the complete ${\cal O}(\epsilon^2)$
lagrangian
in the small ``scale expansion'' one has to add the leading 1/M
$N\Delta$-structures
of chapter 4.3 to this lagrangian.

Finally, we list the corresponding $\Delta\Delta$ form
\begin{eqnarray}
{\cal L}_{\Delta\Delta}^{(2)} &=& \bar{T}^{\mu}_i \left[ a_1 \chi_{+}^{(s)}
+ a_2 \left( v \cdot u \right)^2 + a_3 \; u \cdot u +a_4 \left[
S_\alpha , S_\beta \right] u^\alpha u^\beta  \right. \nonumber \\
& & \phantom{\bar{T}^{\mu}_i [} \left. + a_5 \left( \chi_+ -
\chi_{+}^{(s)} \right) \right] g_{\mu\nu} \; \delta^{ij} \; T^{\nu}_j
\; + i \; \bar{T}^{\mu}_i \left[ a_6 \; f_{\mu\nu}^+ +
a_7 \; v_{\mu\nu}^{(s)} \right] \delta^{ij} \; T^{\nu}_j \; +
\nonumber \\
& &+ \bar{T}^{\mu}_i \left[ a_8 \left\{ w_{\alpha}^a , w_{\beta}^b \right\}
      \left(\delta^{ia}\delta^{jb}+\delta^{ib}\delta^{ja}\right) g_{\mu\nu}
      g^{\alpha\beta} \right] T^{\nu}_j \nonumber \\
& &+ \bar{T}^{\mu}_i \left[ a_9 \left\{ w_{\alpha}^a , w_{\beta}^b \right\}
      \left(\delta^{ia}\delta^{jb}+\delta^{ib}\delta^{ja}\right) g_{\mu\nu}
      v^\alpha v^\beta \right] T^{\nu}_j \nonumber \\
& &+ \bar{T}^{\mu}_i \left[ a_{10} \left\{ w_{\alpha}^a , w_{\beta}^b \right\}
      \left(\delta^{ia}\delta^{jb}+\delta^{ib}\delta^{ja}\right) g_{\mu}^\alpha
      g_{\nu}^\beta \right] T^{\nu}_j \nonumber \\
& &+ \bar{T}^{\mu}_i \left[ a_{11} \left\{ w_{\alpha}^a , w_{\beta}^b \right\}
      \delta^{ij} \delta^{ab} g_{\mu}^\alpha
      g_{\nu}^\beta \right] T^{\nu}_j \; .
\label{eq:ctDD}
\end{eqnarray}
Once more we note that one has to add the corresponding 1/M
spin 3/2 lagrangian of chapter 4.2 to Eq.(\ref{eq:ctDD}) in
order to obtain the O($\epsilon^2$) delta-delta lagrangian which is both
reparameterization invariant and constitutes the proper
non-relativistic limit of the relativistic nucleon-delta-pion system.
The physics content of the first seven terms is straightforward and is in
general identical to the corresponding $NN$ case described above.
The remaining 4 structures start out as $\Delta\Delta\pi\pi$ vertices.

\section{Conclusions}

The subject of chiral perturbation theory in the baryon sector has by
now become highly developed.  Indeed predictions have been made and in
many cases experimentally confirmed for a variety of processes
involving $\pi-\gamma-N$ interactions, as summarized in ref. \cite{j}.
In this work, however, effects of the $\Delta$(1232) are included only in
terms of contributions to the various counterterms which contribute to
the various reactions.  Because of this treatment, any such
predictions in channels to which the $\Delta$(1232) can contribute are
necessarily restricted to the very-near threshold region.  Indeed the
$\Delta N$ mass difference is of the same order of magnitude as the ``small''
parameter $m_\pi$ and cannot be neglected in studies of non-threshold
phenomena.  In this context we have developed a procedure by which the
$\Delta(1232)$ can be treated as an explicit degree of freedom in such
heavy baryon chiral perturbative studies.  The method we have used is
a simple generalization of the familiar nucleon technique of
integrating out the ``heavy'' lower component in favor of the
``light'' upper component, but is more complex due to the presence of
additional spin-1/2 components together with the desired spin 3/2
structure in the Rarita-Schwinger formalism.   Nevertheless, we have
shown how it can be accomplished using heavy baryon projection
operators which isolate the various spin/spinor components of the
wavefunction.  This procedure opens the way then for a rigorous
expansion not just in powers of energy-momentum $p$ and pion mass
$m_\pi$ but also simultaneously in the small quantity
$\Delta_0=(M_\Delta-M_0)$---we call this an expansion in $\epsilon$ as opposed
to
the usual expansion in $p$ and $m_\pi$ which are generically noted by
$p$---and will hopefully allow extension of the near-threshold predictions
given in ref. \cite{j} into higher energy domains.  Herein we have
generated a start to this process by presenting the formalism which makes
such a program possible.  We have shown how, using projection
operators, the heavy spin-3/2 field can be isolated from its
Rarita-Schwinger form and have, in a path integral context,
constructed the lowest order effective action.  We also included
coupling to nucleons and showed how the program can be carried out in
the general case.  Forms for possible counterterm Lagrangians were
presented up to ${\cal O}(\epsilon^2)$.  At this point one can
begin to apply the formalism to physical processes and parameters.
Studies in this regard are under way.

\begin{center}
{\bf Acknowledgments}
\end{center}

It is a pleasure to acknowledge the hospitality of the Institute for
Nuclear theory, where part of this work was done and where important
insights and comments by Malcolm Butler enabled us to get off the
starting blocks. One of us (JK) acknowledges support from the Institut
de Physique Nucl\'eaire, Orsay
\footnote{Laboratoire de Recherche des Universit\'es Paris XI et Paris
VI, associ\'e au CNRS.}
where part of his contribution to this project was performed.
This research is supported in part by the National
Science Foundation, by the Natural Science and Engineering Research
Council of Canada and by the Sweizerischer Nationalfonds.

\newpage

\appendix
\section{The Rarita-Schwinger Formalism }

The free spin 3/2 field of mass $M_{\Delta}$, represented as a vector-spinor
field $\Psi_{\mu} (x)$, satisfies the equation of motion
\begin{equation}
(i \gamma_{\nu} \partial^{\nu} - M_{\Delta} ) \Psi_{\mu} (x) = 0
\end{equation}
with the subsidiary condition
\begin{equation}
\gamma_{\mu} \Psi^{\mu} (x) = 0 \label{eq:subsidiary}
\end{equation}
Given these two equations, one can also show
\begin{equation}
\partial_{\mu} \Psi^{\mu} (x) = 0 \label{eq:subsidiary2}
\end{equation}

We now expand the spin 3/2 field into plane wave states of definite spin
$s_{\Delta} = -\frac{3}{2} ... +\frac{3}{2}$ and momentum $p$ to give an
explicit representation of $\Psi_{\mu} (x)$ in terms of (anti)-particle
creation and annihilation operators $b, b^{\dag}$, $(d, d^{\dag})$ respectively
\begin{equation}
\Psi_{\mu} (x) = \sum_{ s_{\Delta} } \int \frac{d^{3} p }{ J_{F} } \left(
                 b ( {\bf p},s_{\Delta} ) \; {\em u}_{\mu} ( {\bf p},
                 s_{\Delta} ) e^{-i p \cdot x} +
                 d^{\dag} ( {\bf p},s_{\Delta} ) \; {\em v}_{\mu} ( {\bf p},
                 s_{\Delta} ) e^{i p \cdot x} \right)
\end{equation}
where ${\em u}_{\mu} ({\bf p},s_{\Delta})$ is called a Rarita-Schwinger
spinor. For the energy dependent normalization constant $J_{F}$ we choose
\begin{equation}
J_{F}= ( 2 \pi )^{3} \frac{E}{M_\Delta} \; .
\end{equation}
The Rarita-Schwinger spinor for the spin 3/2 field is constructed by coupling
a spin 1 vector $e_{\mu} ({\bf p},\lambda )$ to a spin 1/2 Dirac spinor ${\em
u} ({\bf p},s) $ via Clebsch-Gordon coefficients and then boosting
to a velocity ${\bf v}={\bf p}/M_{\Delta}$
\begin{equation}
{\em u}_{\mu} ({\bf p},s_{\Delta}) = \sum_{\lambda , s} \; (1 \lambda
\frac{1}{2} s | \frac{3}{2} s_{\Delta} ) \; e_{\mu} ({\bf p},\lambda ) \;
{\em u} ({\bf p},s) \label{eq:RSspinor}
\end{equation}
where \cite{EW}
\begin{eqnarray}
e^{\mu} ({\bf p},\lambda ) & = & \left( \frac{ {\bf \hat{e}}_{\lambda} \cdot
                                 {\bf p} }{ M_\Delta } \; , \; {\bf \hat{e}}_{
                                 \lambda} + \frac{{\bf p ( \hat{e}_{\lambda}
                                 \cdot p ) }}{ M_{\Delta} ( p_{0} + M_{\Delta}
)
                                 } \right) \label{eq:spin1} \\
{\em u} ({\bf p},s)       & = & \sqrt{\frac{E + M_{\Delta}}{2 M_\Delta}}
                                \left( \begin{array}{c}
                                 \chi_{s} \\ \frac{{\bf \sigma \cdot p}}{
                                 E + M_{\Delta} } \chi_{s} \end{array} \right)
\end{eqnarray}
For the unit vectors ${\bf \hat{e}_{\lambda}} \; , \lambda = 0, \pm 1$
appearing
in Eq. \ref{eq:RSspinor} we use a spherical representation
\begin{equation}
{\bf \hat{e}}_{+} =  - \frac{1}{\sqrt{2}} \left( \begin{array}{c} 1 \\ i \\ 0
                        \end{array} \right) \hspace{5mm}
{\bf \hat{e}}_{0} = \left( \begin{array}{c} 0 \\ 0 \\ 1 \end{array} \right)
                        \hspace{5mm}
{\bf \hat{e}}_{-} = \frac{1}{\sqrt{2}} \left( \begin{array}{c} 1 \\ - i \\ 0
                        \end{array} \right)
\end{equation}
The anti-particle spinors ${\em v}_{\mu} ({\bf p}, s_{\Delta} )$
can be constructed analogously.

It is important to note that the spin 3/2 field, due to its construction via a
direct spin 1 - spin 1/2 coupling, always contains spurious
spin 1/2 degrees of
freedom. It is therefore useful to introduce a complete set
of orthonormal spin projection operators, which enable separation of the spin
3/2 and spin 1/2 components:
\begin{eqnarray}
\left( P^{3/2} \right)_{\mu \nu} + \left( P^{1/2}_{11} \right)_{\mu \nu} &+&
\left( P^{1/2}_{22} \right)_{\mu \nu}  =  g_{\mu \nu} \\
\left( P^{I}_{ij} \right)_{\mu \delta} {\left( P^{J}_{kl}
\right)^{\delta}}_\nu
& = & \delta^{I J} \delta_{jk} \left( P^{J}_{i l}
\right)_{\mu\nu}
\end{eqnarray}
with
\begin{eqnarray}
\left( P^{3/2}      \right)_{\mu \nu}  &=&  g_{\mu \nu} - \frac{1}{3}
                                            \gamma_{\mu} \gamma_{\nu} - \frac{
                                            1}{3 p^{2}} \left( \gamma \cdot p
\;
                                            \gamma_{\mu} p_{\nu} + p_{\mu}
                                            \gamma_{\nu} \gamma \cdot p \;
                                            \right) \nonumber\\
\left( P^{1/2}_{11} \right)_{\mu \nu} & = & \frac{1}{3} \gamma_{\mu}
                                            \gamma_{\nu} - \frac{p_{\mu}
                                            p_{\nu}}{p^{2}} + \frac{
                                            1}{3 p^{2}} \left( \gamma \cdot p
\;
                                            \gamma_{\mu} p_{\nu} + p_{\mu}
                                            \gamma_{\nu} \gamma \cdot p \;
                                            \right) \nonumber\\
\left( P^{1/2}_{22} \right)_{\mu \nu} & = & \frac{p_{\mu}
p_{\nu}}{p^{2}}
\nonumber\\
\left( P^{1/2}_{12} \right)_{\mu \nu} & = & \frac{1}{ \sqrt{3} p^{2} } \left(
                                            p_{\mu} p_{\nu} - \gamma \cdot p \;
                                            p_{\nu} \gamma_{\mu} \right)
                                            \nonumber\\
\left( P^{1/2}_{21} \right)_{\mu \nu} & = & \frac{1}{ \sqrt{3} p^{2} } \left(
                                            \gamma \cdot p \; p_{\mu}
                                            \gamma_{\nu} - p_{\mu} p_{\nu}
                                            \right) \label{eq:zz}
\end{eqnarray}
Note that there exist {\it two}
spin 1/2 degrees of freedom in addition to the desired spin 3/2 component.
Finally,
we also give the following useful properties of the spin projection operators:
\begin{eqnarray}
\left[ \not\!{p} \; , \left( P^{3/2} \right)_{\mu \nu} \right]_{-}  & = & 0 \\
\left\{ \not\!{p} \; , \left( P^{1/2}_{i j} \right)_{\mu \nu} \right\}_{+}
& = & 2 \; \delta^{i j} \left( P^{1/2}_{i j} \right)_{\mu \nu} \not\!{p}
\end{eqnarray}

\newpage

\section{Isospurion Formalism for $\Delta$(1232) }

$\Delta$(1232) is an isospin 3/2 system. The four physical states $\Delta^{++},
\Delta^{+}, \Delta^{0}, \Delta^{-}$ can be described by treating the spin 3/2
field $\Psi_{\mu} (x)$ as an isospin-doublet and attaching an additional
isovector index $i = 1,2,3$ to it. The resulting field, $\Psi_{\mu}^{i} (x)$,
is therefore a vector-spinor field both in spin and in isospin space. Note that
the vector-spinor construction in isospin space would allow for six states, we
therefore introduce a subsidiary condition, analogously to
Eq.(\ref{eq:subsidiary}), to eliminate two degrees of freedom
\begin{equation}
\tau^{i} \; \Psi_{\mu}^{i} (x) = 0
\end{equation}
where $\tau^{i}$ represents the three Pauli matrices.

For the three isospin doublets we use the representation
\begin{eqnarray}
\Psi_{\mu}^{1} & = & \frac{1}{\sqrt{2}} \left[ \begin{array}{c} \Delta^{++} -
                     \frac{1}{\sqrt{3}} \Delta^{0} \\ \frac{1}{\sqrt{3}}
                     \Delta^{+} - \Delta^{-} \end{array} \right]_{\mu}
\nonumber\\
\Psi_{\mu}^{2} & = & \frac{i}{\sqrt{2}} \left[ \begin{array}{c} \Delta^{++} +
                     \frac{1}{\sqrt{3}} \Delta^{0} \\ \frac{1}{\sqrt{3}}
                     \Delta^{+} + \Delta^{-} \end{array} \right]_{\mu}
\nonumber\\
\Psi_{\mu}^{3} & = & - \sqrt{\frac{2}{3}} \left[ \begin{array}{c} \Delta^{+}
                     \\ \Delta^{0} \end{array} \right]_{\mu} \label{eq:rr}
\end{eqnarray}

As in Appendix A, one can construct a complete set of orthonormal isospin
projection operators \cite{EW}, which we use extensively, to separate the
isospin 3/2 from the isospin 1/2 components
\begin{eqnarray}
\xi_{i j}^{3/2} + \xi_{i j}^{1/2} & = & \delta^{i j } \\
\xi_{i j}^{I} \xi_{j k}^{J}       & = & \delta^{I J} \xi_{i k}^{J}
\end{eqnarray}
with
\begin{eqnarray}
\xi_{i j}^{3/2} & = & \delta^{i j} - \frac{1}{3} \tau^{i} \tau^{j} = \frac{2}
                      {3} \delta^{i j} - \frac{i}{3} \epsilon_{i j k} \tau^{k}
                      \\
\xi_{i j}^{1/2} & = & \frac{1}{3} \tau^{i} \tau^{j} = \frac{1}{3} \delta^{i j}
                      + \frac{i}{3} \epsilon_{i j k} \tau^{k}
\end{eqnarray}

\newpage

\section{Inverses of Matrices ${\bf C_{\Delta}}$ at ${\cal O}({\bf 1/M})$}

At ${\cal O}(1/M)$, matrices ${\cal C}_{\Delta}^{-1}$ (section 2) and
${\bf C}_{\Delta}^{-1}$ (section 3) exhibit a block-diagonal
substructure with block matrices $\tilde{X}^{-1}$, $\tilde{Y}^{-1}$
and $\tilde{Z}^{-1}$, as displayed in Table 2. At this order they read
\begin{eqnarray}
\tilde{X}^{-1}_{\mu \nu} & = & \frac{-1}{2 m} \; P_{-} \;\;
                                    _{(33)}P^{3/2}_{\mu \nu} \; P_{-}
                                    \nonumber\\
                             &   & \nonumber \\
\tilde{Y}^{-1}_{\mu \nu} & = & \frac{-1}{6 m} \left[
                                     \begin{array}{cc}
                                     - 3 P_{+} \; _{(11)}P^{1/2}_{\mu \nu}
                                     \; P_{+} & - \sqrt{3}
                                     P_{+} \; _{(12)}P^{1/2}_{\mu \nu}
                                     \; P_{-} \\
                                     - \sqrt{3}
                                     P_{-} \; _{(21)}P^{1/2}_{\mu \nu}
                                     \; P_{+} & 3 \; P_{-} \;
                                     _{(22)}P^{1/2}_{\mu \nu} \; P_{-}
                                     \end{array}
                                     \right] \nonumber\\
                             &   & \nonumber \\
                             &   & \nonumber \\
\tilde{Z}^{-1}_{\mu \nu} & = & \frac{-1}{6 m} \left[
                                     \begin{array}{cc}
                                     3 \; P_{-} \; _{(11)}P^{1/2}_{\mu \nu}
                                     \; P_{-} & - 3 \; \sqrt{3} \;
                                     P_{-} \; _{(12)}P^{1/2}_{\mu \nu}
                                     \; P_{+} \\
                                     - 3 \; \sqrt{3} \; P_{+} \;
                                     _{(21)}P^{1/2}_{\mu \nu} \; P_{-} &
                                     5 \; P_{+} \; _{(22)}P^{1/2}_{\mu \nu}
                                     \; P_{+}
                                     \end{array}
                                    \right], \label{eq:Z} \\
                             &    & \nonumber
\end{eqnarray}
with $m = M_{\Delta}$ for the pure spin 3/2 case (section 3) and $m = M_{0}$
for the case of simultaneous nucleon and delta degrees of freedom (section 4).

At this point we want to remind the reader concerning
the nature of the $1/M$ expansion
in the latter case. We use $M_{0}$ and $\Delta=M_{\Delta}-M_{0}$ as the two
independent mass scales in the theory and work in the limit $\Delta << M_{0}$
in order to establish a systematic chiral power counting. We have therefore
truncated ${\bf C}_{\Delta}^{-1}$ at ${\cal O}(1/M)$ in order to arrive at
Eq.(\ref{eq:Z}). It happens to be the case that there is no explicit
$\Delta$ dependence in ${\bf C}_{\Delta}^{-1}$ at leading order. However,
this will not be true at ${\cal O}(1/M^{2})$, the matrices
${\cal C}_{\Delta}^{-1}$
and ${\bf C}_{\Delta}^{-1}$ will then start to look very different.
\begin{table}
\hspace{3.5cm}
\begin{tabular}{r|ccccc}
   & 3-            & 1+            & 2-            & 1-            & 2+ \\
\hline
3- & $x_{11}^{-1}$ & 0             & 0             & 0             & 0 \\
1+ & 0             & $y_{11}^{-1}$ & $y_{12}^{-1}$ & 0             & 0 \\
2- & 0             & $y_{21}^{-1}$ & $y_{22}^{-1}$ & 0             & 0 \\
1- & 0             & 0             & 0             & $z_{11}^{-1}$ &
$z_{12}^
                                                                      {-1}$ \\
2+ & 0             & 0             & 0             & $z_{21}^{-1}$ & $z_{22}^
                                                                      {-1}$
\end{tabular}
\caption{Block substructure of matrix $C_{\Delta}^{-1}$ at ${\cal O}(1/M)$}
\end{table}

\end{document}